\documentclass[fleqn,usenatbib]{mnras}

\usepackage{graphicx}
\usepackage{ifthen}
\usepackage{url}
\usepackage{amsmath}
\usepackage{bm}
\usepackage{lscape}
\usepackage{rotating}
\usepackage[graphicx]{realboxes}
\usepackage{supertabular}
\usepackage{pdfpages}
\usepackage{footnote}
\usepackage{hyperref}
\usepackage{amssymb}
\makesavenoteenv{tabular}
\makesavenoteenv{table}
\usepackage[T1]{fontenc}
\usepackage{tipa}
\usepackage{accents}
\newcommand{\ubar}[1]{\underaccent{\bar}{#1}}
\usepackage{tablefootnote}

\hypersetup{
  colorlinks   = true,    
  urlcolor     = blue,    
  linkcolor    = blue,    
  citecolor    = blue      
}

\usepackage{ulem}

\def\ltsima{$\; \buildrel < \over \sim \;$}
\def\lta{\lower.5ex\hbox{\ltsima}}
\def\gtsima{$\; \buildrel > \over \sim \;$}
\def\simgt{\lower.5ex\hbox{\gtsima}}
\def\kms{{\rm\,km \  s^{-1}}}

\def\kpc{{\rm\,kpc}}

\def\msun{{\rm\,M_\odot}}

\usepackage{amsmath}

\def\s{\ifmmode \widetilde \else \~\fi}
\def\={\overline}

\def\spose#1{\hbox to 0pt{#1\hss}}

\def\eg{{e.g.,\ }}
\def\ie{{i.e.,\ }}
\def\lta{\mathrel{\spose{\lower 3pt\hbox{$\mathchar"218$}}
     \raise 2.0pt\hbox{$\mathchar"13C$}}}
\def\gta{\mathrel{\spose{\lower 3pt\hbox{$\mathchar"218$}}
     \raise 2.0pt\hbox{$\mathchar"13E$}}}
\def\Dt{\spose{\raise 1.5ex\hbox{\hskip3pt$\mathchar"201$}}}    
\def\dt{\spose{\raise 1.0ex\hbox{\hskip2pt$\mathchar"201$}}}    

\def\dotsfill{\leaders\hbox to 1em{\hss.\hss}\hfill}

\def\FeH{{\rm[Fe/H]}}

\def\ione{{~\sc i}}
\def\ii{{~\sc ii}}

\newcommand{\teff}{T$_{\rm eff}$}
\newcommand{\logg}{log\,$g$}

\loadboldmathitalic 
\title[GHOST P1836849]{
Probing the early Milky Way with GHOST spectra of an extremely metal-poor star in the Galactic disk } 
\author[A. Dovgal et al.] 
{Anya Dovgal$^{1,2}$,
Kim A.\,Venn$^{1}$\thanks{Email: \url{kvenn@uvic.ca}},
Federico Sestito$^{1}$,
Christian R.\,Hayes$^{3}$,
Alan W.\,McConnachie$^{3,1}$, 
\newauthor
Julio F.\,Navarro$^{1}$,
Vinicius M.\,Placco$^{4}$,
Else Starkenburg$^{5}$,
Nicolas F.\,Martin$^{6,7}$,
John S.\,Pazder$^{3,1}$,
\newauthor 
Kristin Chiboucas$^{8}$,
Emily Deibert$^{8}$,
Roberto Gamen$^{9}$,
Jeong-Eun Heo$^{10}$,
Venu M. Kalari$^{10}$,
\newauthor
Eder Martioli$^{11}$,
Siyi Xu$^{8}$,
Ruben Diaz$^{10}$,
Manuel Gomez-Jimenez$^{10}$,
David Henderson$^{8}$,
\newauthor
Pablo Prado$^{10}$,
Carlos Quiroz$^{10}$,
J.\,Gordon Robertson$^{15,16}$,
Roque Ruiz-Carmona$^{10}$,
Chris Simpson$^{8}$,
\newauthor
Cristian Urrutia$^{10}$,
Fletcher Waller$^{1,10}$,
Trystyn Berg$^{3,12}$,
Gregory Burley$^{3}$,
Zachary Hartman$^{8}$,
\newauthor
Michael Ireland$^{13}$,
Steve Margheim$^{14}$,
Gabriel Perez$^{10}$,
Joanna Thomas-Osip$^{10}$
\\
\\
$^{1}$ Department of Physics and Astronomy, University of Victoria, PO Box 3055, STN CSC, Victoria BC V8W 3P6, Canada\\
$^{2}$ Department of Physics and Astronomy, University of British Columbia, 6224 Agricultural Road, Vancouver, BC V6T 1Z1, Canada\\
$^{3}$ NRC Herzberg Astronomy \& Astrophysics, 5071 West Saanich Road, Victoria, BC V9E 2E7, Canada\\
$^{4}$ NSF’s NOIRLab, Tucson, AZ 85719, USA\\
$^{5}$ Kapteyn Astronomical Institute, University of Groningen, Landleven 12, NL-9747AD Groningen, the Netherlands\\
$^{6}$ Universit\'e de Strasbourg, CNRS, Observatoire astronomique de Strasbourg, UMR 7550, F-67000 Strasbourg, France\\
$^{7}$ Max-Planck-Institut f\"{u}r Astronomie, K\"{o}nigstuhl 17, D-69117 Heidelberg, Germany \\
$^{8}$ Gemini Observatory/NSF’s NOIRLab, 670 N. A’ohoku Place, Hilo, HI, 96720, USA \\
$^{9}$ Instituto de Astrof\'isica de La Plata, CONICET--UNLP, Paseo del Bosque s/n, 1900, La Plata, Argentina \\
$^{10}$ Gemini Observatory/NSF’s NOIRLab, Casilla 603, La Serena, Chile \\
$^{11}$ Laborat\'orio Nacional de Astrof\'isica, Rua Estados Unidos 154, 37504-364, Itajub\'a, MG, Brazil \\
$^{12}$ Dipartimento di Fisica G. Occhialini, Università degli Studi di Milano Bicocca, Piazza della Scienza 3, I-20126 Milano, Italy\\
$^{13}$ Research School of Astronomy and Astrophysics, Australian National University, Canberra 2611, Australia \\
$^{14}$ Vera C. Rubin Observatory, NSF's NOIRLab, Casilla 603, La Serena, Chile\\
$^{15}$ Australian Astronomical Optics, Macquarie University, 105 Delhi Rd, North Ryde NSW 2113, Australia\\
$^{16}$ School of Physics, University of Sydney, NSW 2006, Australia
}

\pubyear{2023}
\date{Accepted XXX. Received YYY; in original form ZZZ}
\begin{document}
\maketitle 
\label{firstpage}
\pagerange{\pageref{firstpage}--\pageref{lastpage}}

\begin{abstract}
Pristine\_183.6849+04.8619
(P1836849) is an extremely metal-poor ([Fe/H]$=-3.3\pm0.1$) star on a prograde orbit confined to the Galactic disk. 
Such stars are rare and may have their origins in protogalactic fragments that formed the early Milky Way, in low mass satellites accreted later, or forming in situ in the Galactic plane.
Here we present a chemo-dynamical analysis of the spectral features between $3700-11000$ \AA\ from a high-resolution spectrum taken during Science Verification of the new Gemini High-resolution Optical SpecTrograph (GHOST).
Spectral features for many chemical elements are analysed (Mg, Al, Si, Ca, Sc, Ti, Cr, Mn, Fe, Ni), and valuable upper limits are determined for others (C, Na, Sr, Ba).
This main sequence star exhibits several rare chemical signatures, including 
(i) extremely low metallicity for a star in the Galactic disk, 
(ii) very low abundances of the light $\alpha$-elements (Na, Mg, Si) compared to other metal-poor stars, and 
(iii) unusually large abundances of Cr and Mn, where [Cr, Mn/Fe]$_{\rm NLTE}>+0.5$.
A comparison to theoretical yields from supernova models 
suggests that two low mass Population III objects (one 10 M$_\odot$ supernova and one 17 M$_\odot$ hypernova) can reproduce the abundance pattern well (reduced $\chi^2<1$).
When this star is compared to other extremely metal-poor stars on quasi-circular, prograde planar orbits, differences in both chemistry and kinematics imply there is little evidence for a common origin. The unique chemistry of P1836849 is discussed in terms of the earliest stages in the formation of the Milky Way.  
\end{abstract}

\begin{keywords}
Galaxy: formation - Galaxy: evolution - stars: abundances - stars: kinematics and dynamics - stars: Population III
\end{keywords}

\section{Introduction}

Low-metallicity stars are among the oldest stars in the Galaxy. Cosmological simulations suggest that these pristine stars formed within 2--3 Gyr after the Big Bang, preferentially in low-mass protogalactic systems \citep[\eg][]{Starkenburg17a, ElBadry18,Sestito21}. As the Milky Way (MW) grows, these protogalactic systems contribute their stars, gas, and dark matter contents throughout the proto-MW, including some into planar orbits  that will later form the disk \citep{Sestito21,Santistevan21}. 
Low-mass systems accreted later are expected to disperse their stars primarily into the halo 
\citep{Bullock2005, Johnston2008}, though simulations show that they can also contribute stars with nearly circular orbits on the Galactic disk  \citep{Abadi03,Scannapieco11,Sestito21, Santistevan21}. Contributions to the disk are also possible from an in-situ component of stars formed from the deposited gas \citep{Abadi03,Navarro2018,Yu2021}, and even the chaotic pre-disk epochs when stars are born in irregular configurations \citep[\eg][]{Belokurov22, Belokurov23}. 
Some simulations suggest the transition from "bursty" to "steady" star formation occurs after a stable hot gaseous halo surrounds the MW progenitor, 
impacting the gas accretion mechanisms, such that a coherent disk forms  
via dissipative accretion,
i.e., when the angular momentum of the accreting gas is aligned with the forming galaxy disk \citep[e.g.,][]{Sales2012,Stern2021,Hafen2022}.
This later formation means that "steady" star formation in the disk would occur from pre-enriched gas.

Nevertheless, some extremely metal-poor (EMP, [Fe/H]$<-3$) stars have been found confined to the Galactic plane \citep[\eg][]{Sestito19, Sestito20, Venn20, Kielty21, FernandezAlvar2021, Cordoni21}. It is not clear if these stars occupy the extreme metal-poor extension of the thin disk or the high rotating tail of hotter MW structures like the thick disk or the halo.
A comparison of EMP stars with planar orbits on prograde vs retrograde orbits does show a net 
preference for prograde stars 
in both observations \citep{Sestito20} and simulations \citep{Sestito21,Santistevan21}.
If true, it could suggest an additional source of prograde EMP stars compared to retrograde stars, which are almost certainly accreted from protogalactic fragments and low mass satellites during the early Galactic assembly.    
For example, it is possible that some quasi-circular prograde planar EMP stars may have formed in situ at very early times in the formation of the Galactic proto-disk.
Alternatively, a dwarf galaxy whose orbit was brought into the disk and circularized before being tidally disrupted could have added its stars to the proto-Galactic disk.

While dynamics alone might not help us to clearly identify planar stars that formed in situ, chemo-dynamical analyses can provide more clues.  The most chemically pristine stars in the Milky Way are expected to have been enriched by only one or a few Population~III (Pop~III) supernovae or hypernovae events \citep[\eg][]{Frebel10,Heger12,Ishigaki18}.  
Recently, the aluminum abundance in metal-poor stars has been proposed as a way to disentangle stars that formed in situ from those accreted from satellites  \citep[\eg][]{Das20,Belokurov22}.
However this indicator is limited to stars with [Fe/H]$>-2$.  At lower metallicities, differences in the [Al/Fe] (and most other light element ratios) are less distinct between different stellar populations 
\citep[e.g.,][]{Yong13, Yong21, Aoki13, Skuladottir21}. 
Below [Fe/H]=$-3$, it has been suggested that neutron-capture elements may differ between EMP stars in ultra-faint dwarf (UFD) galaxies when compared with similar stars in the Galactic halo and classical dwarf galaxies \citep[e.g.,][]{Jablonka15, Ji19, Sitnova21}, particularly the [Sr/Ba] ratios. 
If EMP stars have been enriched by a very small number of supernovae, then the ultimate goal would be to use this information to trace their origins back to their host.  
This is work in progress, as nucleosynthetic yields and our understanding of galaxy formation and early star formation improve.

Currently, only seven EMP stars with quasi-circular prograde planar orbits have had their detailed chemical abundances analysed;
SDSS~J102915$+$172927 \citep{Caffau11, Caffau12},
2MASS~J1808202$-$5104378 \citep{Schlaufman2018, Mardini22b},
four stars from the SkyMapper survey (when orbits from \citealt{Cordoni21} are cross matching with abundances from \citealt{Yong21}), and Pristine\_183.6849+04.8619 (P1836849, \citealt{Venn20}).
P1836849 was discovered as part of the spectroscopic follow-up studies to the Pristine survey \citep{Starkenburg17b}, 
a narrow-band imaging survey using MegaCam at the Canada-France-Hawaii Telescope.   Using a specialized Ca\ii{} HK filter in combination with broad-band photometry, the Pristine survey has demonstrated high efficiency in the detection of metal-poor stars  \citep[i.e., $>$56 percent accuracy at $\FeH\leq-2.5$,][]{Youakim17, Aguado19, Martin2023}. 
P1836849 was noted as an EMP with a quasi-circular prograde orbit \citep{Venn20}, which, as discussed above, is uncommon amongst EMP stars. 
%

In this paper, we present a new orbital and chemical analysis of P1836849.  New spectra were taken during the System Verification observations of the new Gemini High-resolution Optical SpecTrograph \citep[GHOST,][]{Pazder20} at Gemini South as described in Section~\ref{sec:data}.  
GHOST is the ideal instrument as it has very high efficiency and wide spectral coverage (3700 - 11000 \AA), making it possible to estimate precise abundances for a large number of elements.  The potential of GHOST spectroscopy has been made clear by the analysis of two stars in the Reticulum II dwarf galaxy \citep{Hayes23} and one metal-poor star in the Milky Way that was either accreted from a low mass satellite or formed in one of the low-mass building blocks of the proto-Galaxy \citep{Sestito23gh}. 
The determination of new orbital and stellar parameters are described in Section~\ref{sec:stellparam}. 
Our spectral line analyses and chemical abundance determinations from model atmospheres are described in Sections~\ref{sec:analysis} and \ref{sec:chems}.  
A comparison to other prograde EMP stars in the disk, EMP stars in the MW halo and nearby low-mass galaxies, and theoretical nucleosynthetic yields are presented in Section~\ref{sec:discussion}, as the basis for our discussion on the origins of this star.   
Overall, we note that the higher efficiency and larger wavelength coverage of GHOST makes this an excellent instrument for the determination of precision chemical abundances and radial velocities 
for stars in the Local Group (i.e.,  G$\lesssim$19, dependent on the signal-to-noise requirements).

\begin{table*}
\caption{
The long and short names of the EMP quasi-circular planar stars, and their Gaia DR3 source IDs and photometric indices (G and BP$-$RP). Reddening (A$_{\rm V}$) for the first three stars is from \citet{Schlafly11} and used to calculate our heliocentric distances (see text).  $^*$Reddening and distances for the last four stars are taken from \citet{Cordoni21}.} 
\label{tab:target}
\begin{tabular}{lllcccccc}
\hline
Target  & Short name & source ID   & G & BP$-$RP & A$_{\rm V}$ &  D  \\
& & & (mag) & (mag) &  (mag)  & (kpc)  \\
\hline
Pristine\_183.6849+04.8619   & P1836849 & 3894267325687326592 & 14.82  & 0.62 & 0.05  &   $1.20\pm0.12$   \\

SDSS J102915+172927 & SDSS J102915 & 3890626773968983296
 & 16.53 & 0.79 & 0.07  & $1.49\pm0.32$ \\

2MASS J18082002-5104378 & 2MASS J18082002 &  6702907209758894848
& 11.75 &  0.91 & 0.31 & $0.59\pm0.01$ \\

SMSS J133308.90-465407.9 & SMSS J133308 &  6083921475163462528
&   12.22     &  1.28   & 0.10 &  \,\, $3.83\pm0.96^*$     \\

SMSS J190556.70-454724.2 & SMSS J190556 & 6710975288644549760
& 12.90       &  1.22   & 0.07 & \,\, $6.19\pm1.54^*$ \\

SMSS J190836.24-401623.5 & SMSS J190836 & 6717349947823371776
& 13.10       &  1.27   & 0.10 & \,\, $5.62\pm0.66^*$ \\

SMSS J232121.57-160505.4 & SMSS J232121 & 2406023396270909440
& 12.53       &  0.95   & 0.02 & \,\, $1.10\pm0.20^*$ \\

\hline
\end{tabular}
\end{table*}

\section{Data}\label{sec:data}

\subsection{Target Selection}

During early testing and calibration of the Pristine survey, spectroscopic observations of bright (V$<15$) metal-poor candidates were observed with the Canada-France-Hawaii Telescope's ESPaDoNS high resolution spectrograph \citep{Donati2006}.  
Out of 115 metal-poor candidates analysed by \citet{Venn20},
one target, Pristine\_183.6849+04.861 (P1836849), was found to have an unusual chemical and kinematic behaviour. 
Orbital analysis using Gaia DR2 \citep{Gaia16,GaiaDR2} showed that P1836849 has a quasi-circular orbit (eccentricity $\epsilon \sim 0.3$) with a relatively small maximum height from the MW plane (Z$_{\rm max}$ = 1.2 kpc). The object is close to the Sun (distance 
$\sim 1.05$ kpc) and to its apocentre (R$_{\rm apo} = 8.4$ kpc; see Table~\ref{tab:params}).
Despite its low metallicity ([Fe/H]$=-3.25$; see Section~\ref{sec:analysis}), this
star was unlikely to be a MW halo interloper because of its  
low Mg and low Na abundances ([Mg/Fe] $= 0.13$, [Na/Fe] $= -0.18$; 
although these were NLTE corrected abundances, and consistent with the NLTE corrected abundances of other EMP MW halo stars examined by \citealt{Venn20}).
At the time, it was proposed as an accreted object early in the Galaxy's lifetime.   
Here, we update the analysis of P1836849, starting with its basic information from Gaia DR3 \citep{GaiaDR3} provided in Table~\ref{tab:target}.

\begin{table}
\caption{GHOST science and calibration exposures for P1836849 (Program ID: GS-2023A-SV-101). These observations were taken in the standard resolution, single object mode, with 2x2 binning.} 
\label{tab:exp}
\centering
\hspace{-0.6cm}
\begin{tabular}{llrcr}
\hline
Filetype & Arm & t$_{\rm exp}$ & N$_{\rm exp}$ & SNR @$\lambda$ \\
      &     & (s) & & (\AA) \\
\hline
Science \\
GS-2023A-SV-101-19-001 &
Blue     &  1800 & x2 &  60 @4130 \\
 &  Red  & 1200 & x3 &  90 @6500 \\
Calibrations \\
GS-CAL20230511-15-001 & arc & 720 & -- & -- \\
GS-CAL20230511-14-001 & flat & 30 & -- & --\\
GS-CAL20230511-20-001 & 2x2 bias & -- & -- & --\\
GS-CAL20230510-1-001  & 1x1 bias & -- & -- & --\\


\hline
\end{tabular}
\end{table}

\begin{figure*}
\includegraphics[width=1\textwidth]{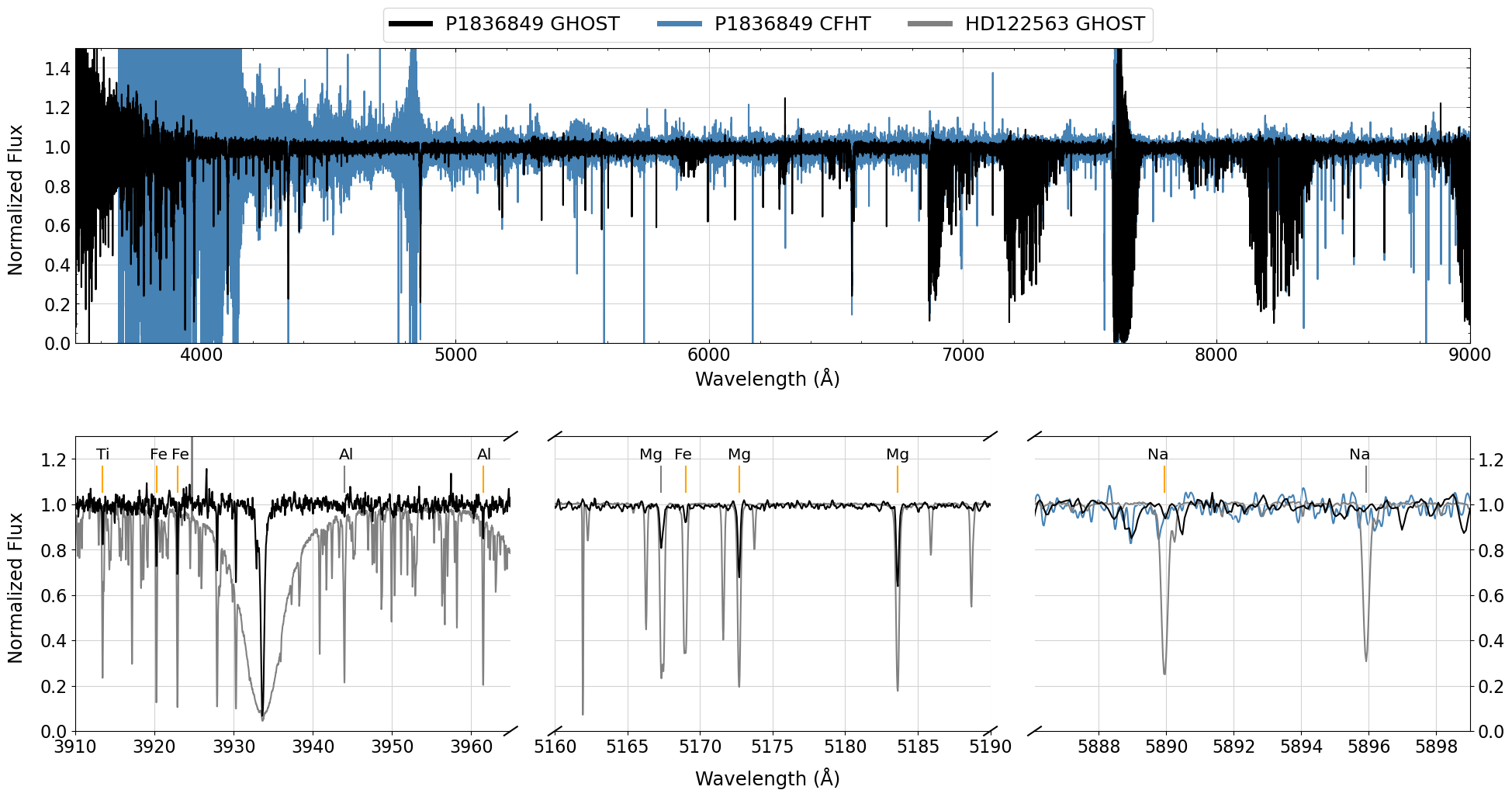}
\caption{Top panel: Spectrum of P1836849 from Gemini-GHOST (black, [Fe/H]$=-3.3$) is compared to the CFHT-ESPaDoNS observation (blue), 
over a wide spectral region from $3500-9000$ \AA.
Lower panel: Three spectral windows show \ion{Ti}{II}, \ion{Fe}{I}, and \ion{Al}{I} lines from $3910-3965$ \AA{}, the \ion{Mg}{I} and \ion{Fe}{I} lines from $5150-5200$ \AA{}, and the absence of the \ion{Na}{D} lines in P1836849 from $5885-5900$ \AA{}. A GHOST spectrum for HD\,122563  (grey, note [Fe/H]$=-2.8$) is shown for comparison in the lower panels. 
We also show the CFHT spectrum for P1836849 near the \ion{Na}{D} lines only to emphasize the weakness of these features.
Spectral lines used in our chemical analysis are as marked in orange.}
\label{Fig:spectra}
\end{figure*}

\subsection{GHOST observations} 

P1836849 was observed on 10 May 2023\footnote{We thank the Gemini Systems Verification team for this birthday gift for 2 out of the first 3 co-authors.}, during the System Verification run of the new GHOST spectrograph \citep{Ireland2012, McConnachie2022}.  
The instrument setup chosen was the standard resolution mode (SR: $R\sim 50,000$) and target mode IFU1:Target—IFU2:Sky. Each IFU in this SR mode includes 7 hexagonal fibers in a compact arrangement projected to 1.2 arcseconds on-sky.  These fibres are then aligned to form a pseudo-slit which enters the spectrograph, delivering light to two cameras (red and blue).  This design means no light losses at the slit edges, thereby delivering all the light within 1.2" to the spectrograph.  The nominal wavelength coverage of the two cameras is $360-542$ nm (blue) and $517-1000$ nm (red), however some light is transmitted beyond these boundaries but with rapidly decreasing quantum efficiencies.
As seen in Table~\ref{tab:exp}, multiple exposures were taken with 2x2 binning (i.e., CCD binning in the spatial and spectral directions, and in both the red and blue channels).  These exposures were taken at a mean air mass AM = 1.49, and the Moon was $\sim75\%$ illuminated.

The spectra were processed using the GHOST Data Reduction pipeline v1.0 (\textsc{GHOST DR} - originally described by \citealt{Ireland18} and \citealt{Hayes22}), which was modified by the \textsc{DRAGONS} \citep{DRAGONS2022} team during the commissioning of GHOST.  
\textsc{DRAGONS} (Data Reduction for Astronomy from Gemini Observatory North and South) is a Python-based, open-source platform for the reduction and processing of astronomical data at Gemini.
With the calibration files also listed in Table~\ref{tab:exp}, this pipeline completed all the steps for the reduction of spectroscopic data from 2D CCD images
(i.e., bias/flat corrections, wavelength calibration, sky subtraction, barycentric correction, extraction of individual orders, and variance-weighted stitching of the spectral orders), including the more complicated file management inherent to GHOST observations.
The GHOST DR delivered 1D spectra for each of the blue (3) and red (5) exposures.
For each region, we combined the exposures using the median flux, then normalized using asymmetric k-sigma clipping (over 10 \AA\ regions).  Unfortunately, one of the blue and two of the red exposures appeared to have no flux, reducing the number of blue exposures to 2 (from 3) and red exposures to 3 (from 5).  
In a final step,
the combined blue and red spectra were further combined with a weighted average in the (small, $517-542$ nm) overlapping region, and taking into account the variance of each spectrum. 

The final spectrum was corrected for radial velocity offsets, determined using \textsc{iraf/fxcor} \citep{Tody86,Tody93} and a template spectrum of the EMP standard star HD122563 (from GHOST; see \citealt{Hayes23}). For this step, the spectral region for RV fitting was reduced to 3900 - 6600 \AA\ to avoid increasing noise at the shortest wavelengths and variations in the telluric features at longer wavelengths.  The heliocentric corrected radial velocity for P1836849 from GHOST is RV~$= 38.9\pm0.1\kms$, which is in good agreement (2$\sigma$) with RV~$= 40.0\pm0.5\kms$
from the CFHT spectrum.

Compared to the CFHT spectrum, the GHOST spectrum has equal or much higher SNR at all wavelengths; see Figure~\ref{Fig:spectra} (top panel). 
The CFHT spectrum\footnote{CFHT data archives, including the spectrum of P1836849, at \\ www.cadc-ccda.hia-iha.nrc-cnrc.gc.ca/en/cfht/}
is comprised of 2x 2400~s exposures designed for SNR~$\sim30$ near the \ion{Mg}{I} b 5170 \AA, whereas the GHOST spectrum includes only 2x 1800~s exposures in the same spectral region for SNR~$\sim90$.
The bottom panels in Figure~\ref{Fig:spectra} compare the GHOST spectrum of our main sequence star P1836849 to the GHOST spectrum of the EMP standard red giant star HD122563 ([Fe/H]~$=-2.8$, \teff~$=4642$ K, logg~$=1.26$; \citealt{Hayes23}) in three regions. 
It seems that P1836849 is more metal-poor than HD122563, however its atmosphere is also warmer and denser which weakens and broadens the spectral lines independent of metallicity. 
In the bottom right panel of Figure~\ref{Fig:spectra}, we also include the CFHT spectrum for P1836849 (blue) to emphasize that the \ion{Na}{I D} lines are not present,
but provide valuable upper limits (see Section~\ref{sec:chems}). 
The apparent noise in the GHOST spectrum in this region is due to imperfect telluric line removal (partially due to weather conditions and partially due to the air mass for this $\sim$equatorial target).

For our GHOST spectrum of P1836849, 
the SNR values range from (25: 60: 90) near (3800: 4100: 6500 \AA). 
Most importantly, the GHOST spectrum extends very blueward (to 3700 \AA), which allows us to reach important spectral features such as \ion{Al}{I} 3961 \AA, \ion{Eu}{II} 4129 \AA, \ion{Sr}{II} 4077 and 4215 \AA, \ion{Ba}{II} 4554 \AA, and CH 4300 \AA.  
Even if these spectral lines are not detected, they can provide valuable upper limits useful for chemo-dynamical analyses of EMP stars.

\section{Orbital and Stellar Parameters}\label{sec:stellparam}

Distance and stellar parameters have been updated from those reported in \citet{Venn20} using Gaia DR3 \citep{GaiaDR3} and improvements in our methodology described here.

\subsection{Astrometric distance}
The astrometric distances of P1836849, SDSS~J102915, and  2MASS~J18082002 are derived using their exquisite Gaia DR3 parallaxes in a Bayesian framework. The posterior probability distribution function is obtained multiplying a Gaussian likelihood on the parallax, shifted by the zero-point offset \citep{Lindegren21}, and a Galactic halo stellar density distribution prior \citep[see][for further details]{Sestito19}.

\begin{table*}
\caption{Stellar and orbital parameters. Stellar parameter uncertainties from this paper are discussed in Section~\ref{sec:stellarparameters}. 
The references are 
(a) This work, 
(b) \citet{Venn20}, 
(c) \citet{Caffau12}, 
(d) \citet{Sestito19}, 
(e) \citet{Mardini22b},
(f) \citet{Schlaufman2018}, (g) \citet{Cordoni21} and \citet{Yong21}.
}
\label{tab:params}
\centering
\resizebox{\textwidth}{!}{

\begin{tabular}{lccccccccr}
\hline
Star & T$_{\rm eff}$  & log(g) & [Fe/H] &  RV &  R$_{\rm apo}$ & R$_{\rm peri}$ & Z$_{\rm max}$ & ecc & Ref  \\
& (K) &  &  &  ($\kms$)  & (kpc) & (kpc) & (kpc) & & \\
\hline
P1836849 & $6478\pm247$ & $4.41\pm0.11$ & $-3.25\pm0.11$ & $38.5\pm0.1$ & $8.39\pm0.04$ & $3.75\pm0.23$ & $1.28\pm0.10$ & $0.38\pm0.03$  & a  \\
   & $6491\pm42$ & $4.44\pm0.03$ & $-3.16\pm0.07$ &  $40.0\pm0.5$ & $8.5\pm0.1$ & $4.5\pm0.1$ & $1.2\pm0.1$ & $0.30\pm0.01$ & b \\
\hline
SDSS & $5784\pm189$ & $4.76\pm0.23$ & -- & -- & $8.76\pm0.21$ & $7.32\pm0.11$ & $2.36\pm0.60$ & $0.09\pm0.02$ & a \\
J102915   & $5800\pm75$ & $3.9\pm0.3$ & $-4.89\pm0.06$ & $-34.5\pm0.1$ & -- & -- & -- & -- & c \\  
 & $5764\pm57$  & $4.7\pm0.1$ &  -- &  -- &  $10.93\pm0.23$ & $8.62\pm0.05$  & $2.37\pm0.23$ & $0.12\pm0.01$ & d \\

\hline
2MASS & $5630\pm143$ & $3.46\pm0.06$ & -- & -- & $7.52\pm0.01$ & $4.80\pm0.04$ & $0.13\pm 0.01$ & $0.22\pm0.01$ & a \\
J18082002   & $5665$ & $3.34$  & $-3.85$ & $16.7\pm0.1$ & 7.8 & 5.2  &  0.6 & $0.22\pm0.01$ & e \\  
  & $6124\pm44$ & $3.5\pm0.1$ & --  & -- & $7.60\pm0.05$ & $6.33\pm0.11$  & $0.165\pm0.006$ & $0.09\pm0.01$ & d \\
 & $5400\pm100$ & $3.0\pm0.2$ & $-4.07\pm0.07$  & $16.5\pm0.1$ & $7.66\pm0.02$ & $5.56\pm0.07$ & $0.126\pm0.004$ & $0.158\pm0.005$ & f  \\

\hline
SMSS J133308 & -- & -- & -- & -- & $6.40\pm0.18$ & $3.95\pm0.36$ & $1.69\pm0.53$ & $0.24\pm0.03$  & a  \\
   & $4900\pm100$ & $1.75\pm0.3$ & $-2.79\pm0.3$ &  $-16.0\pm0.8$ & $6.57^{+0.24}_{-0.08}$ & $5.03^{+0.53}_{-0.17}$ & $1.58^{+0.59}_{-0.48}$ & $0.13^{+0.01}_{-0.03}$ & g \\

\hline
SMSS J190556 & -- & -- & -- & -- & $3.44\pm0.62$ & $2.09\pm0.55$ & $2.31\pm0.64$ & $0.23\pm0.06$  & a  \\
   & $4850\pm100$ & $1.62\pm0.3$ & $-2.73\pm0.3$ &  $-45\pm2.3$ & $4.61^{+0.89}_{-0.38}$ & $2.95^{+0.89}_{-1.27}$ & $2.77^{+0.62}_{-0.72}$ & $0.22^{+0.21}_{-0.03}$ & g \\

\hline
SMSS J190836 & -- & -- & -- & -- & $3.67\pm0.29$ & $2.92\pm0.34$ & $2.06\pm0.31$ & $0.12\pm0.03$  & a  \\
   & $4825\pm100$ & $1.51\pm0.3$ & $-3.33\pm0.3$ &  $-44.2\pm1.1$ & $5.93^{+0.20}_{-0.07}$ & $3.05^{+0.51}_{-0.58}$ & $2.91^{+0.63}_{-0.45}$ & $0.29^{+0.11}_{-0.07}$ & g \\

\hline
SMSS J232121 & -- & -- & -- & -- & $9.49\pm0.43$ & $4.99\pm0.31$ & $1.38\pm0.05$ & $0.31\pm0.05$  & a  \\
   & $5450\pm100$ & $3.23\pm0.3$ & $-3.03\pm0.3$ &  $-39.1\pm1.0$ & $10.38^{+0.55}_{-0.62}$ & $5.86^{+0.46}_{-0.35}$ & $1.44^{+0.29}_{-0.30}$ & $0.28^{+0.05}_{-0.06}$ & g \\

\hline
\end{tabular}

}
\end{table*}


The new heliocentric distance for P1836849  is within $1.08\sigma$ of the distance determined using Gaia DR2 data  \citep{Venn20}, however the older distance used a more complex Bayesian method that combined astrometric and photometric data with an extremely metal-poor set of MESA/MIST isochrones \citep{Choi16,Dotter16}, a prior on the Galactic stellar density distribution, and a prior on the age of the metal-poor stars to give a probability distribution function on the  distance, as fully described in \citealt{Sestito19}. By excluding the use of isochrones with the high precision Gaia DR3 parallax, then we can avoid  systematics in the poorly constrained metal-poor isochrones \citep[e.g.,][]{Heiter2015,Karovicova20}.

The Gaia DR3 parallax and the new derived heliocentric distances are reported in Table~\ref{tab:target}.

\begin{figure}
\centering
\includegraphics[width=0.44\textwidth]{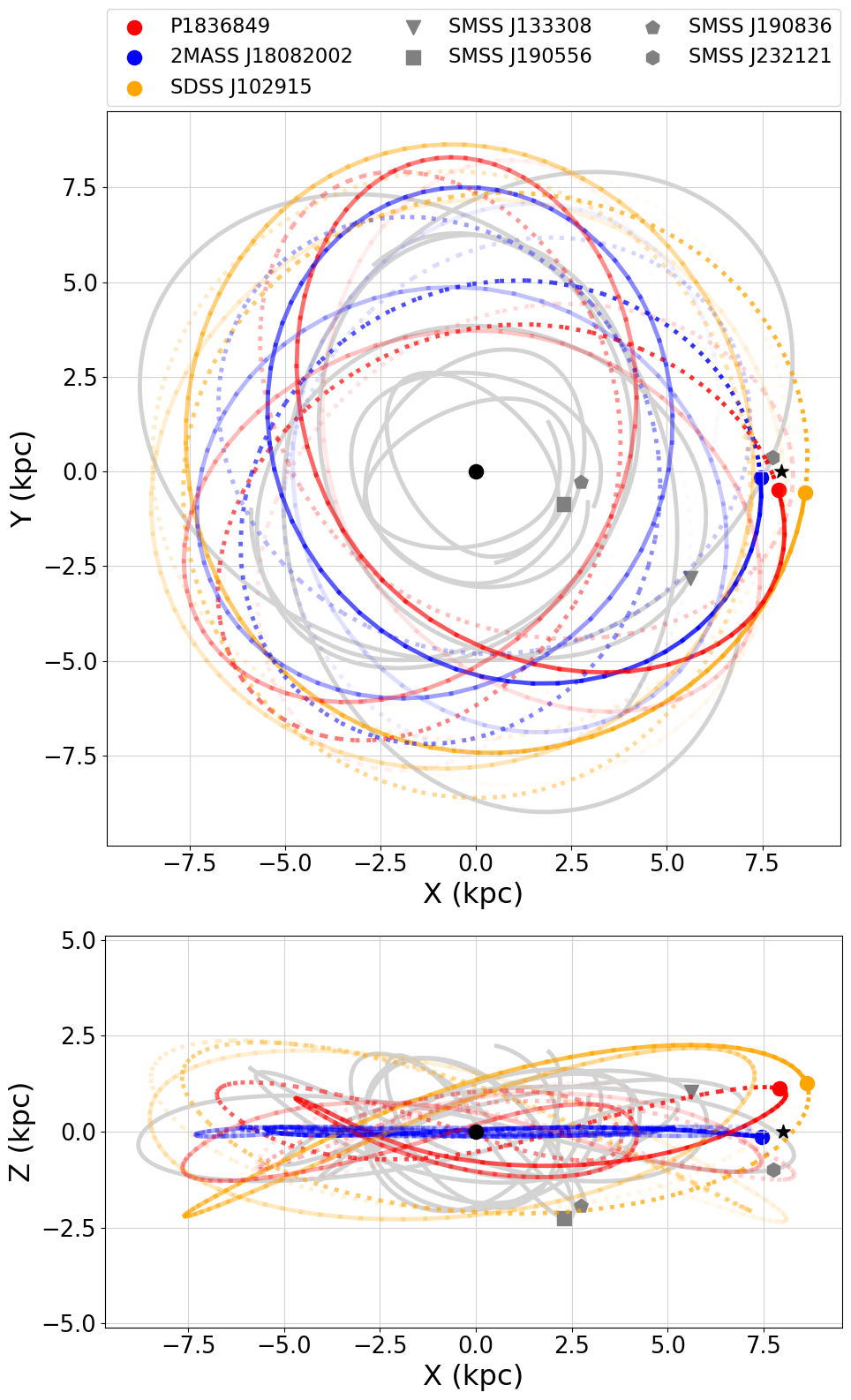}
\caption{Galactic orbital motion. Top panel: Galactic Y vs. X. Bottom panel: Galactic Z vs. X. The positions at the present time of P1836849, SDSS~J102915, and 2MASS~J18082002 are marked by the red, orange, and blue circles, respectively. Solid and dotted lines of similar colour denote the orbits integrated backwards and forwards.
The orbits for the four SkyMapper stars are shown in grey (solid lines only, though their orbits are also integrated backwards and forwards). Their current positions are noted as grey symbols.
Black circle and black star mark the position of the Galactic centre and of the Sun. }
\label{fig:orbits}
\end{figure}

\subsection{Orbital parameters}
The 6D kinematic data for P1836849 has been updated from Gaia DR2 to Gaia DR3 values, including the new astrometric distance, and  the new RV determined from the GHOST spectrum (see Table~\ref{tab:params}).
Orbital parameters are derived using \textsc{galpy} \citep{Bovy15}, where the same Galactic gravitational potential as in \citet{Sestito19} and \citet{Venn20} has been adopted. This briefly consists in the \textsc{MWPotential2014} with an increased dark matter halo mass of $1.2 \times 10^{12}\rm M_{\odot}$ \citep{BlandHawthorn16}. Uncertainties on the orbital parameters are derived from a Monte Carlo simulation on the input parameters (distance, RV, proper motion, coordinates), drawing them from  Gaussian distributions for 1000 times. Then, the median and the standard deviation are used to represent the measurement of a parameter and its uncertainty.
Present and previous orbital parameters are listed in Table~\ref{tab:params}.
The apocentric distance and the maximum height from the plane are in agreement within less than $1\sigma$ from the previous measurements. The new pericentric distance is smaller than previously inferred, resulting in a slightly higher eccentricity. In both cases, the star has a prograde motion.

As a comparison, the orbital parameters of SDSS~J102915 and of 2MASS~J18082002 are also re-derived using Gaia DR3 and the methods described above; summarised in Table~\ref{tab:params}. These two stars were found in a very similar kinematical configuration as of our target \citep{Schlaufman2018,Sestito19,Mardini22b}. The updated apocentric and pericentric distances of SDSS~J102915 are now smaller than previously inferred with Gaia DR2 \citep{Sestito19}, and these updates have only small effects on the orbit eccentricity and its maximum height from the plane. The updates for 2MASS~J18082002 result in a smaller pericenter and larger 
eccentricity  \citep{Schlaufman2018,Sestito19}, also seen by \citet{Mardini22b}, and its orbit 
has a remarkably small maximum height from the plane ($\sim0.13\kpc$).
The new Galactic orbits for the three stars are shown in Figure~\ref{fig:orbits}, integrating forwards and backwards by 0.5 Gyr each.

Finally, we add the four SkyMapper stars to Figure~\ref{fig:orbits}, using their Gaia DR3 positions and proper motions in our potential, with distances and radial velocities from \citet{Cordoni21}. 
For clarity, all four orbits are marked with grey solid lines, but their orbits are also integrated backwards and forwards by 0.1-0.2 Gyr each (for clarity).  Only one of the SkyMapper stars is near the main-sequence (SMSS J232121, a sub-giant), placing it in the solar neighbourhood at present, similar to P1836849, SDSS J102915, and 2MASS J18082002. 

\subsection{Stellar parameters}
\label{sec:stellarparameters}

A first estimate of the effective temperature (\teff) 
for P1836849 was 
determined using the colour-temperature relation for Gaia photometry from \citet{Mucciarelli21}. This calibration was selected based on their inclusion of very metal-poor stars \citep[from][]{GonzalezHernandez2009} and has been very successful when applied to the analyses of extremely metal-poor stars \citep[\eg][]{Kielty21, Waller23, Sestito23}. 
A first estimate of the surface gravity (log~$g$) is determined using the Stefan-Boltzmann equation
\citep[\eg see][]{Venn2017,Kraft2003} and assuming the first estimate on \teff. 
These estimates were iterated several times for a convergence on the final \teff{} and log~$g$ \citep[see][for a full description]{Sestito23}. Uncertainties are derived with a Monte Carlo simulation, drawing all the input parameters (distance, G, BP$-$RP, A$_V$, [Fe/H] - as well as the correlated uncertainties of \teff\ and logg)  from a Gaussian distribution for $10^{5}$ times. A flat mass distribution between 0.5 to 0.8 $\msun$ is assumed in the surface gravity uncertainty. A 10\% uncertainty in extinction 
is adopted throughout.

Stellar parameters and uncertainties for P1836849 are reported in Table~\ref{tab:params}. 
These new stellar parameters are within the $1\sigma$ errors of the previous estimates by \citealt{Venn20}. However, the \textit{uncertainties} on the effective temperature are larger compared to the previous estimates based on isochrones. This is a concomitance of two effects. The first is that methodologies based on isochrones can underestimate the intrinsic systematic errors in the theoretical models. The second is due to the photometric temperature calibration itself, where \citet{Sestito23} showed that the large uncertainty only occurs for the hotter stars in the upper main sequence and the sub-giant branch. 
Microturbulence ($\xi=1.3\kms$) was adopted from the calibrations for metal-poor dwarfs by \citep{Sitnova2015}.

Effective temperatures and surface gravities were also re-determined for SDSS~J102915 and 2MASS~J18082002. For SDSS~J102915, the updated stellar parameters are in agreement with the inference based method using Gaia DR2 \citep{Sestito19}, confirming the star is a dwarf. For 2MASS~J18082002, the temperature is now in agreement with the values from \citet{Mardini22b} and \citet{Schlaufman2018}, while the surface gravity confirms its sub-giant nature.

\section{Spectral Line Analyses} \label{sec:analysis}

Chemical abundances 
in P1836849 were
determined from individual spectral lines.  Spectral lines were selected from the recent GRACES and ESPaDoNS analyses of metal-poor halo stars \citep{Venn20, Kielty21, Lucchesi22}, and updated with a search of the P1836849 GHOST spectrum for additional lines from spectrum syntheses (described below).  All atomic data and additional spectral lines were taken from the recent version of {\sl linemake}\footnote{ Available at \url{https://github.com/vmplacco/linemake}} atomic and molecular line database \citep{Placco21}, see Tables~\ref{tab:lines_fe} and \ref{tab:lines_others}. 
We note that hyperfine structure (HFS) components were only significant for our results for two spectral lines: 
\ion{Sc}{II} 4246.822 \AA{} and 
\ion{Mn}{I} 4030.746 \AA.  

Chemical abundances have been determined from a classical model atmospheres analysis using the stellar parameters in Table~\ref{tab:params}.  Model atmospheres are 
from  \textsc{MARCS}\footnote{\url{https://marcs.astro.uu.se}} \citep{Gustafsson08}, and we restrict the analysis to relatively unblended and weak spectral lines (\ie equivalent width EW $<130$ m\AA).
Chemical abundances are compared to the Sun using standard notation\footnote{[X/Y] = log n(X)/n(Y)$_*$ $-$ log n(X)/n(Y)$_\odot$, where n(X) and n(Y) are column densities (in cm$^{-2}$)}, and solar abundances from \citet{Asplund09}.

\subsection{Spectrum Syntheses}
\label{syntheses}

The 1D LTE radiative transfer code \textsc{MOOG}\footnote{MOOG (Nov 2019 version) is available at \url{http://www.as.utexas.edu/~chris/moog.html}} \citep[][]{Sneden73, Sobeck11} was used to synthesise the stellar spectra using the stellar parameters as described above.
This method was carried out in three steps: (1) a model atmosphere was generated with the initial parameters:  \teff, \logg, and $\xi$ as described in Section \ref{sec:stellparam}, and an initial metallicity of [Fe/H]$=-3.2$.
%
The iron lines were synthesised for a preliminary metallicity estimate, and the model atmosphere updated with the new metallicity.  This process was repeated until the metallicity output matched the input (typically only twice). (2) A new synthesis of all elements was generated which included line abundances and upper limits for all of the clean spectral lines.  (3) NLTE (below) and HFS corrections were applied.
Each synthetic spectrum was broadened in \textsc{MOOG} to match the observed spectrum; we found that a Gaussian smoothing kernel with FWHM~$=0.17$ was a good match to the GHOST spectral resolution and internal thermal broadening for this main sequence star. 
If the spectral features were well fit, then we calculated an abundance for that line from the syntheses.  If not, then 
a 3$\sigma$ maximum equivalent width was used to calculate an upper limit on the abundances (i.e., this was applied to Na, Sr, Ba, and Eu).  
This method was also used to synthesise  the CH molecular feature near 4300 \AA\, (see Section~\ref{sect:carbon} for details). 

\subsection{Checking the stellar parameters}\label{sec:checkstellar}

It is possible to check the stellar parameters from spectroscopic features, in particular; a flat distribution of A(Fe\ione) as a function of (i) excitation potential ($\chi$) indicates an appropriate effective temperature, (ii) wavelength indicates appropriate sky subtraction and data reduction, (iii) line strength indicates an appropriate microturbulence value ($\xi$), and (iv) an ionisation balance between Fe\ione{} and Fe\ii{} is typically employed to determine the optimal surface gravity.

Our analysis of P1836849 found a slope d[A(Fe\ione)/$\chi$] $<0.1$ dex eV$^{-1}$, which falls well within 1$\sigma$ of the [\ion{Fe}{I}/H] measurements. A similar result was obtained even after applying NLTE corrections, thus confirming our adopted \teff.  No statistically significant slope was found for A(Fe\ione) vs wavelength or line strengths.  The latter confirms our microturbulence value, which was set from the empirical relation for cool dwarfs from \cite{Sitnova2015} that depends on surface gravity - however, our surface gravity value itself is less certain, as 
[\ion{Fe}{I}/H] = [\ion{Fe}{II}/H] +0.2 (LTE) or +0.3 (NLTE). Recent findings by \citet{Karovicova20} indicate that A(Fe\ione{}) can deviate by as much as +0.7 dex from A(Fe\ii{}) in very metal-poor red giants, but only approximately $+0.1\pm0.1$ dex for EMP dwarf stars, like P1836849. Thus, our offset of +0.2 to +0.3 dex based on only a few A(Fe\ii{}) lines seems reasonable, and 
we refrain from adjusting the surface gravity values any further.  We consider our stellar parameters to be appropriate.

\begin{table}
\caption[]{Averaged LTE and NLTE chemical abundances, and final total uncertainty $\sigma$ (which has been divided by the square root of the number of lines N). 
NLTE corrections are
from the following References: 
(a) \citet{Lind2012}, 
(b) \citet{Lind2011}, 
(c) \citet{Bergemann2017}, 
(d) \citet{Bergemann2013}, 
(e) \citet{Mashonkina17}, 
(f) \citet{Bergemann2011}, 
(g) \citet{Bergemann2010b}, 
(h) \citet{Bergemann2019}, 
(i) \citet{Nordlander17a}. 
*Fe species are [X/H] instead of [X/Fe].

}
\centering
\resizebox{0.47\textwidth}{!}{
\hspace{-0.6cm}
\begin{tabular}{lrccrc}
\hline
Species & [X/Fe] & $\sigma$ \ \   & N   & [X/Fe] & REF \\
        & LTE \      &  &  &      NLTE  & \\
\hline
Fe\ione{}* &  $-3.22$  &  $0.05$ & 39  & $-3.08$ & a \\
Fe\ii{}*   &  $-3.42$  &  $0.11$ & 3   &  $-3.41$ & a \\
CH        &   $<+0.80$ &   $-$   & $-$ &   $-$ &   $-$ \\
Na\ione{} &   $<-0.62$ &  $0.15$ & 3   &  $<-0.80$ & b \\
Mg\ione{} &   $0.05$   &  $0.11$ & 4   &   $0.03$ & c \\
Al\ione{} &   $-0.60$  &  $0.15$ & 1   & $-0.23$ & i  \\
Si\ione{} &   $0.25$   &  $0.15$ & 1   &  $0.22$ & d \\
Ca\ione{} &   $0.05$   &  $0.12$ & 1   &  $0.20$ & e \\
Sc\ii{}   &   $0.24$   &  $0.15$ & 1   &  $-$     & $-$ \\ 
Ti\ii{} &   $0.58$  &  $0.12$ & 7  &  $0.51$ & f \\
Cr\ione{} &   $0.07$  &  $0.12$ & 2  &  $0.50$ & g \\
Mn\ione{} &   $-0.08$  &  $0.15$ &  1 &   $0.41$ & h \\
Ni\ione{} &   $0.30$  &  $0.15$ & 1  &  $-$ & $-$ \\
Sr\ii{} &   $<-0.10$  &  $-$ & 2  &  $-$ & $-$ \\
Ba\ii{} &   $<-0.50$  &  $-$ &  1 &  $-$ & $-$ \\
Eu\ii{} &   $<+3.40$  &  $-$ & 1  &  $-$ & $-$ \\
\hline
\end{tabular}}
\label{tab:chems}
\end{table}

\begin{figure*}
\centering
\includegraphics[width=1\textwidth]{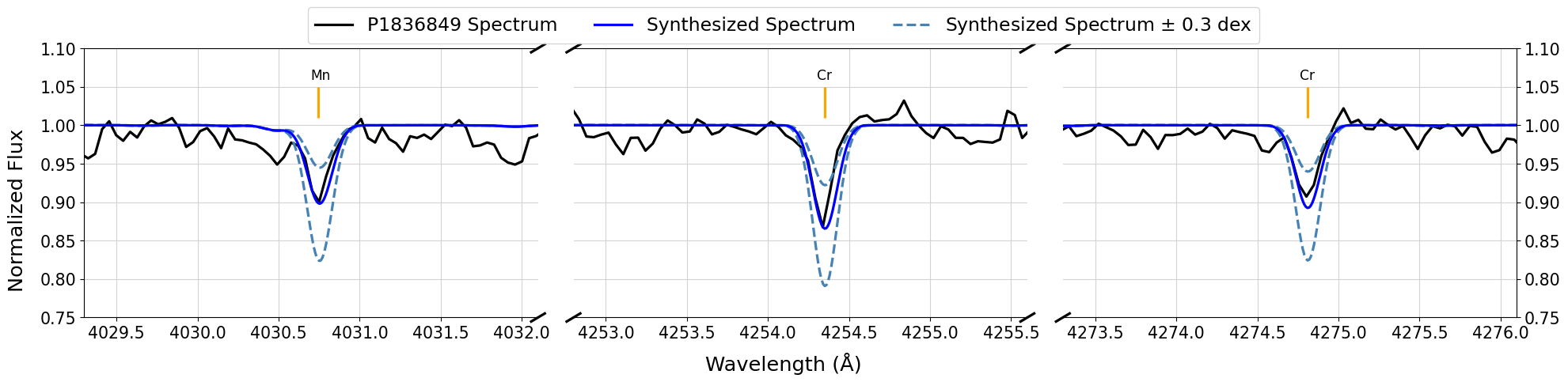}
\includegraphics[width=1\textwidth]
{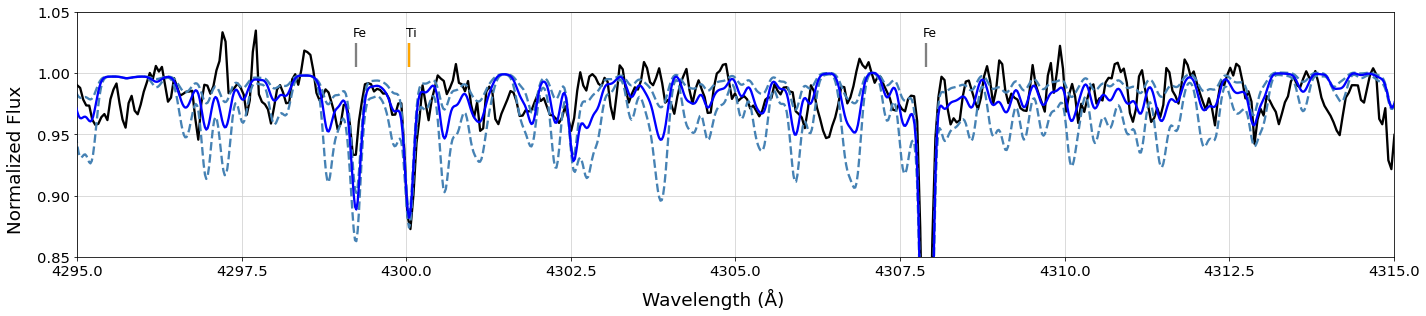}
\caption{Upper Panels: The \ion{Mn}{I} and two \ion{Cr}{I} lines used in this analysis are shown, including our best fit syntheses, and $\Delta$[X/Fe]~$\pm0.3$.  Though the lines are weak ($\lesssim30$ m\AA\ each), they are clear and well modelled. 
Lower Panel: The G-band including our best upper-limit syntheses, with $^{\rm 12}$C/$^{\rm 13}$C$=40$ (EMP dwarf, \eg  \citealt{Spite2021}) and [C/Fe]$=+0.8$. Two more syntheses show $\Delta$[C/Fe]$\pm0.3$. 
One \ion{Ti}{II} line used in this analysis (orange) and two \ion{Fe}{I} lines not used (grey) are also indicated. The plot is zoomed in for clarity. 
}
\label{Fig:cr_mn}
\end{figure*}

\section{Chemical abundances analysis}\label{sec:chems}

The wavelength coverage of GHOST allows us to observe spectral lines of carbon, $\alpha$-, odd-Z, Fe-peak, and neutron-capture process elements.  In total, 61 spectral features are measured in this analysis of P1836849, which is significantly more than the nine lines in total analysed by \cite{Venn20}.  A search for additional clean, unblended spectral lines did not produce any more suitable for an abundance analysis. The chemical abundances and uncertainties are presented in Table~\ref{tab:chems}.

\subsection{Carbon}
\label{sect:carbon}
Carbon was examined from spectrum synthesis of CH near 4300 \AA\ using the updated molecular line list from \cite{Masseron2014} available in \textsc{linemake}. We also adopted $^{\rm 12}$C/$^{\rm 13}$C$=40$  based on the recent finding for the EMP subgiant HD\,140283 \citep{Spite2021}. 
We found no evidence for a carbon enrichment, with an upper-limit of [C/Fe] $< +0.8$; 
see Fig.~\ref{Fig:cr_mn} (note that the wing of H$\gamma$ extends to this region and has been removed in both the observed and synthetic spectra for this plot). 
Examination of the N and O abundances showed negligible effects. Changes in the isotopic ratio of
$\Delta$($^{\rm 12}$C/$^{\rm 13}$C)$\pm10$ resulted in $\Delta$[C/Fe]$=\mp0.1$.
Due to the high temperature and gravity of this star, our non-detection of carbon is primarily due to the feasibility of the line formation itself, and not the SNR of the GHOST spectrum.

\subsection{$\alpha$-elements}
 
The $\alpha$-elements with detectable spectral lines in P1836849 are \ion{Mg}{I} (4), \ion{Si}{I} (1), \ion{Ca}{I} (1), and \ion{Ti}{II} (7).
The A(Mg\ione{}) is an average of the abundances from 2 lines of the Mg\ione{} Triplet ($\lambda\lambda 5172.68, 5183.60$ \AA, the third line is blended with iron and ignored) and 2 lines in the blue spectrum (at 3829.35\,\AA\ and 3832.30\,\AA).  The latter two lines are amongst the strongest lines in our analyses, and \ion{Mg}{I} 3832.30\,\AA\ is in a noisy region of spectrum, yet both have EW~$\lesssim130$\,m\AA\ and are not extremely sensitive to the microturbulence ($\xi$) values, thus we have kept them in our analysis.
Only one line of Si\ione{} is detected at $3905.52$\,\AA. 
Similarly, only the resonance line of Ca\ione{} at $4226.72$\,\AA\ was detected, both sufficiently weak and in a clean spectral region. 
We do not include an analysis for calcium of the strong Ca\ii{} Triplet, as each line has EW~$\gtrsim150$\,m\AA. 
Ti\ii{} was observable from 7 weak spectral lines ranging between $3913.4$ and $4571.9$\,\AA.

\subsection{Odd-Z elements}

The abundances of odd-Z elements have a strong dependence on the metallicities of their progenitors, seen as a strong odd-even effect in low metallicity stars \cite[e.g.,][]{Nomoto13}.  We were able to measure only two spectral lines of odd-Z elements;  Al\ione{} 3961.52 \AA\ and Sc\ii{} 4246.82 \AA{}.  The former is shown in Figure~\ref{Fig:spectra} and the latter spectral feature has hyperfine structure that is included in our spectrum synthesis analysis.  We also examined the Na\ione{} Doublet ($\lambda\lambda 5889.95$ and 5895.92 \AA), but could not clearly detect the lines.  As shown in Figure~\ref{Fig:spectra}, there is significant telluric contamination near the \ion{Na}{D} feature as this target (DEC=+5$^{\rm o}$) was observed through a high airmass at Gemini-South (strong atmospheric bands can also be seen from 6900 to 8400 \AA).  A re-examination of the ESPaDOnS spectrum taken at lower air mass at the CFHT \citep[Northern hemisphere;][]{Venn20} also suggests that the \ion{Na}{D} lines in P1836849 are in the noise, and absent when compared to our standard star HD\,122563.  We use the GHOST spectrum to determine an upper-limit for sodium in P1836849; using both spectrum synthesis and a maximum (3$\sigma$) equivalent width EW = 13 m\AA, we find [Na/Fe]$_{NLTE}<-0.8$. 

\subsection{Fe-peak elements}

The Fe-peak elements observable in our GHOST spectrum include \ion{Fe}{I} (39), \ion{Fe}{II} (3), \ion{Cr}{I} (2), 
\ion{Mn}{I} (1), and \ion{Ni}{I} (1).  This is a significant increase compared to \cite{Venn20} where only 2 lines each of A(Fe\ione) and A(Fe\ii) were available - all 4 were re-analysed in our GHOST spectrum. Our final iron abundance for P1836849 is [Fe/H]$=-3.3\pm0.1$, which is the (unweighted) average of our \ion{Fe}{I} and \ion{Fe}{II} results in Table~\ref{tab:chems}, in both LTE and NLTE.
New Fe-peak spectral lines include the two \ion{Cr}{I} resonance lines detected at $\lambda$4254.35 and $\lambda$4274.81 \AA, and the \ion{Mn}{I} resonance line at $\lambda4030.74$ \AA; these features and our spectrum syntheses are shown in Figure~\ref{Fig:cr_mn}.  We note that the \ion{Mn}{I} exhibits hyperfine structure taken into account in our spectrum synthesis. A weak \ion{Ni}{I} line is also detected at $3858.29$ \AA.

\subsection{Neutron-capture elements}
The high-quality blue spectral coverage of the GHOST spectrograph opens new possibilities for the detection and precision measurements of neutron-capture elements in metal-poor star.  However, our target P1836849 does not include any of the heavy elements as it is too warm and not r-process rich.  We calculate upper-limits only on the abundances of Sr, Ba, and Eu. While our non-detections for the \ion{Sr}{II} 4077.70, 4215.51 \AA, and \ion{Ba}{II} 4554.03 \AA\  resonance lines provide interestingly low upper-limits ([Sr/Fe]~$<-0.1$, [Ba/Fe]~$<-0.5$), the \ion{Eu}{II} 4129 \AA\ upper-limit does not provide a useful constraint ([Eu/Fe]~$<+3.4$; 
hyperfine structure and isotopic components are included in this spectrum synthesis).

\subsection{Non-Local Thermodynamic Equilibrium corrections}

\begin{figure}
\centering
\includegraphics[width=0.47\textwidth]{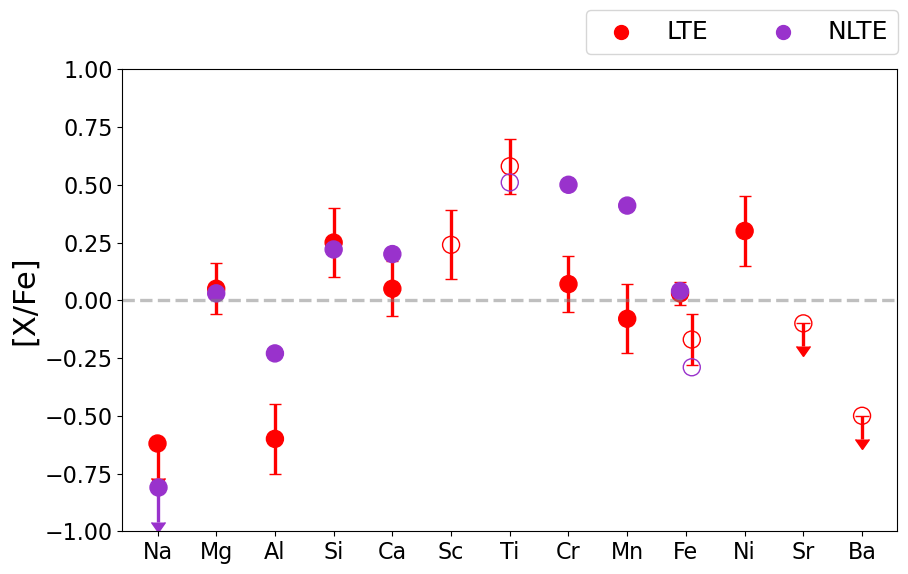}
\caption{P1836849 LTE (red) and NLTE (purple) chemical abundances compared to solar. Filled points are for neutral species, while open circles are for singly-ionized species.  Error bars are shown for LTE abundances only.
}
\label{Fig:abund}
\end{figure}

The radiation field in the atmospheres of EMP stars contributes to significant non-local thermodynamic equilibrium (NLTE) effects, which can be large for some species.
NLTE corrections have been applied whenever possible, 
using corrections tabulated in the MPIA data base\footnote{\url{https://nlte.mpia.de}}. For Na\ione{} and Fe\ii{}, we use corrections available in the \textsc{INSPECT}\footnote{\url{http://www.inspect-stars.com}} database, and for Al\ione{}, we apply NLTE corrections from  \cite{Nordlander17a}.  
References for the NLTE\footnote{NLTE corrections were automized using a new python code for sampling the INSPECT or MPIA databases; available at \url{https://github.com/anyadovgal/NLTE-correction}.}  corrections for individual elements are also in Table~\ref{tab:chems}. 1D LTE and NLTE abundances are shown in Figures~\ref{Fig:abund}~and~\ref{fig:ssolar}.


\subsection{Chemical abundance uncertainties}


In Table~\ref{tab:chems}, we report the chemical abundance ratios from our 1DLTE analysis as [X/Fe]$_{\rm{LTE}}$.
The total error $\sigma_{\rm A(X)}$ includes the effects due to uncertainties in the stellar parameters ($\delta_{\rm T_{eff}}$, $\delta_{\rm logg}$), added in quadrature with the measurement errors.  
Measurement errors due to continuum placement and SNR are computed per line, and combined per species such that $\delta_{\rm X}=\delta_{\lambda}/\sqrt{{\rm N_X}}$.

\section{Discussion}
\label{sec:discussion}

\begin{figure}
\centering
\includegraphics[width=0.47\textwidth]{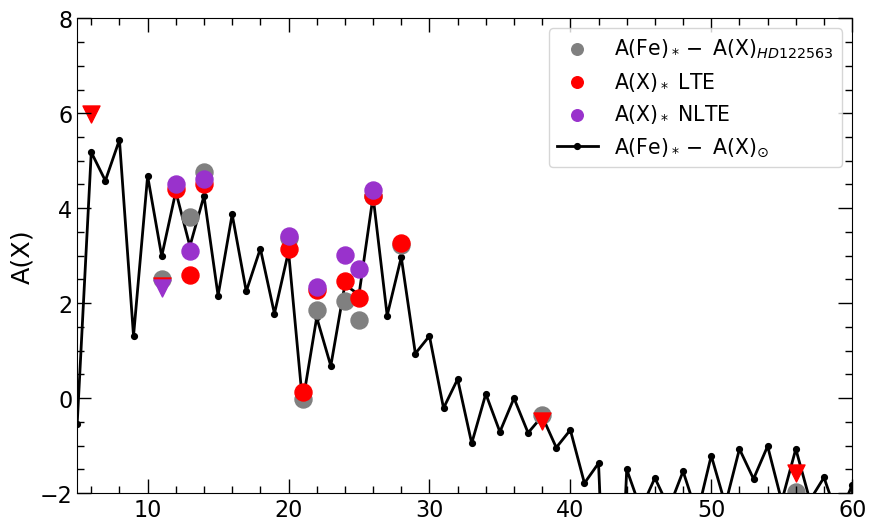}
\caption{P1836849 LTE (red) and NLTE (purple) chemical abundances vs atomic number compared to scaled-solar abundances ($\Delta$[Fe/H]~$=-3.3$), including upper limits (downward triangles) for C, Na, Sr, \& Ba.  
}
\label{fig:ssolar}
\end{figure}

The chemistry of P1836849 is compared directly to solar and scaled-solar abundances (reduced by [Fe/H]$=-3.3$) in Fig.~\ref{Fig:abund} and Fig.~\ref{fig:ssolar}.  
Regardless of whether the LTE or NLTE abundance ratios are examined, P1836849 is not similar to the Sun -  particularly the very low ratios of Na, Al, and Ba, as well as the high ratios of Ti, Ni and the NLTE-corrected values of Cr and Mn.
It is clear that P1836849 formed in a region with a very different star formation history and chemical evolution than that of the Sun.

\subsection{Comparison with MW disk stars }

\begin{figure}
\centering
\includegraphics[width=0.47\textwidth]{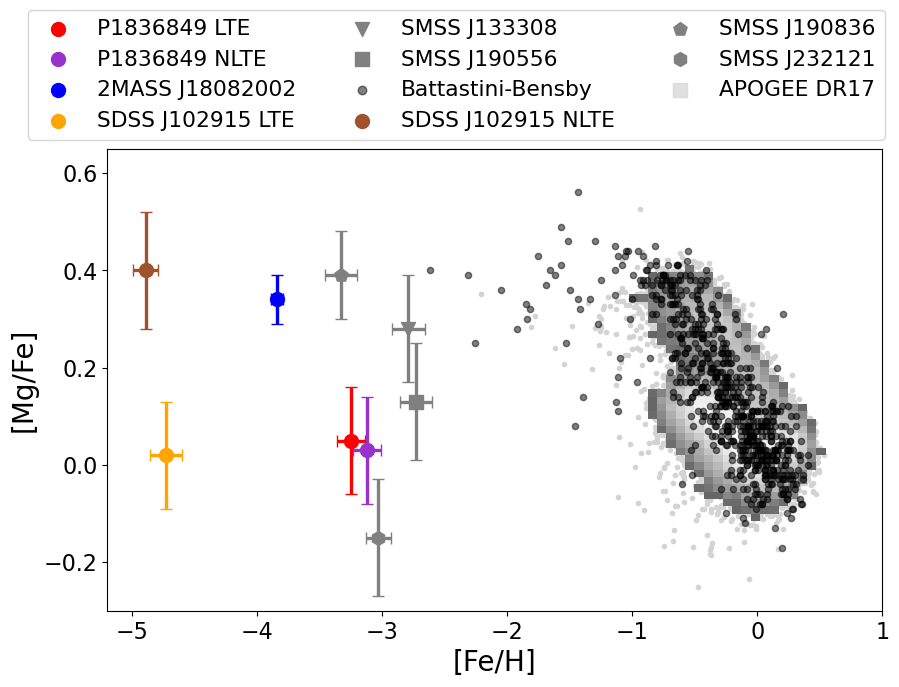}
\caption{LTE [Mg/Fe] vs [Fe/H] for stars in the Galactic disk(s) from \citet{Bensby2014}, \citet{Battistini2015, Battistini2016}, and selected from APOGEE DR17 \citep[][see text]{APOGEEDR17}.  The EMP stars with high resolution spectroscopic abundances and prograde quasi-circular planar orbits from Fig.~\ref{fig:orbits} and Table~\ref{tab:params} are also shown. 
The NLTE abundances for P1836849 and SDSS J102915 are also shown.
}
\label{fig:MWdisk}
\end{figure}

The majority of known stars with planar kinematics have metallicities [Fe/H]$>-2$.  A detailed homogeneous survey of the chemical elements in these stars \citep{Bensby2014, Battistini2015, Battistini2016} describes the chemical enrichment of the MW thin and thick disks in terms of yields from SN II (for stars with [Fe/H]$<-0.4$) and the later contributions from SN Ia and AGB stars.  
In Fig.~\ref{fig:MWdisk}, we show [Mg/Fe] for these disk stars (black markers), as well as those available in the Sloan Digital Sky Survey Data Release 17 \citep{sdssdr17} from the Apache Point Observatory Galactic Evolution Experiment \citep[APOGEE,][]{apogee2017}\footnote{APOGEE data were taken using the SDSS-2.5m telescope \citep{Gunn2006AJ} and the LCO-2.5m Ir\'{e}n\'{e}e du Pont telescope \citep{Bowen1973}, and a description of the APOGEE instruments and data processing can be found in \citet{Wilson2019} and \citet{Nidever2015}, respectively.  
The targeting for the APOGEE survey is described in \citet{Zasowski2013,Zasowski2017,Beaton2021,Santana2021}.  The stellar parameters and chemical abundances for the APOGEE data were measured as described in \citet{GarciaPerez2016} using the linelist described in \citet{Shetrone2015} and \citet{Smith2021}.} (grey markers). The APOGEE stars were selected to have |Z|$<$3 kpc, and space velocities $150<$ V$_\phi<250\kms$, and |V$_{\rm R}$| and |V$_{\rm Z}$| $<30\kms$ \citep[][calculated using APOGEE radial velocities, STARHORSE distances, and Gaia EDR3 proper motions]{Queiroz2020, gaiaedr3}. Clearly, P1836849 does not resemble the stars that describe the MW disk, nor belong to a population extrapolated to very low metallicities of either the thin or thick disk stars.

\subsection{Comparison with EMP planar stars}

We compare the chemistry and kinematics of P1836849 with the six other known EMP stars with prograde quasi-circular planar orbits that currently have detailed chemical abundances from high resolution spectroscopy: see Tables~\ref{tab:target} and \ref{tab:params}.    
Their orbits are shown in Fig.~\ref{fig:orbits}, and, at first glance, look quite similar. However, upon closer examination, 
the eccentricities vary by a factor of $\sim4$, and two of the SkyMapper stars have quite small apocentric distances.
The minimum eccentricity is that of SDSS J102915 ($\epsilon\sim0.09$) and the maximum is that of P1836849 ($\epsilon\sim0.38$)\footnote{The orbital eccentricity for P1836849 was formerly ecc=$0.3$ from Gaia DR2.  This eccentricity was adopted when selected stars from the SkyMapper survey from \citet{Cordoni21}, which was also based on Gaia DR2. }
Furthermore, the orbit of SDSS J102915 reaches a maximum height of $\sim2.2\kpc$, a factor of two larger than P1836849, and much larger than the very flat orbit of 2MASS J18082002.

\begin{figure}
\centering
\includegraphics[width=0.47\textwidth]{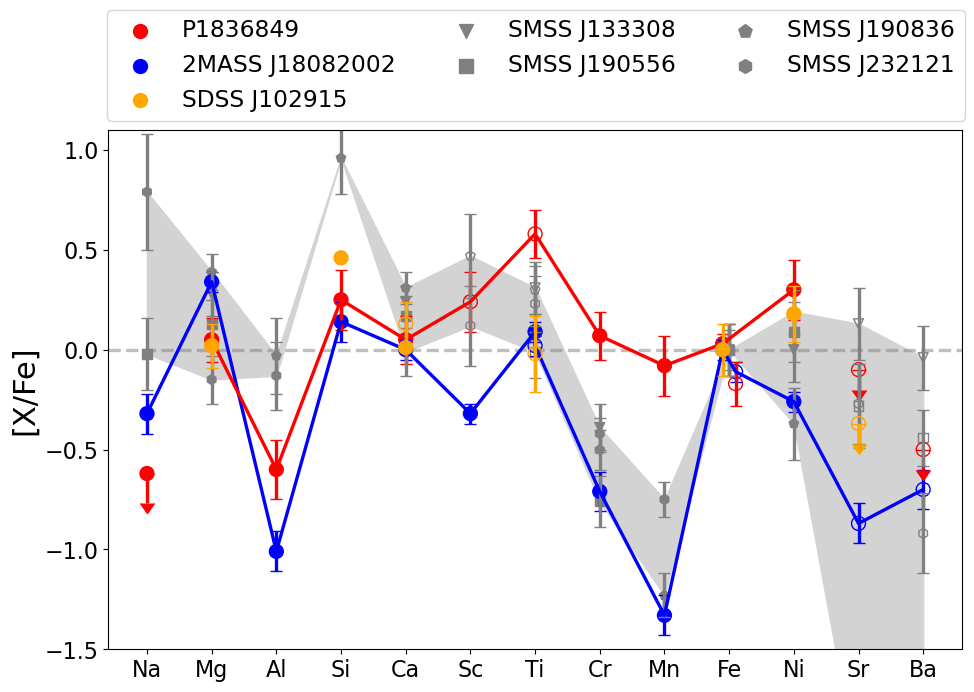}
\caption{A comparison of the LTE stellar abundances in 
P1836849 (red; this paper), 
SDSS J102915 \citep[orange;][]{Caffau11,Caffau12}, 
2MASS J18082002 \citep[blue;][]{Schlaufman2018, Mardini22}, 
and the four SkyMapper stars \citep[gray;][]{Yong21}. Gray shading connects the highest and lowest values amongst the SkyMapper stars, for clarity. P1836849 appears to be chemically distinct from the comparison stars, and a wide range in abundances is seen for the whole sample.
}
\label{Fig:cafabund}
\end{figure}

The LTE abundances of these stars are compared in Fig.~\ref{Fig:cafabund}; LTE abundances are compared as NLTE corrections were not applied in the other analyses.
Unfortunately, the stellar parameters of these seven stars are not very similar; P1836849 is hotter than the comparison stars by $>$800 K, two
comparison stars are subgiants rather than dwarfs, and three of the SkyMapper stars are red giants. 
Furthermore, SDSS J102915 and 2MASS J18082002 are more metal-poor than P1836849 by $\gtrsim1.0$ dex; see Table~\ref{tab:params}. 
These differences impact our ability to directly compare their abundances as systematic errors are not well constrained.
Nevertheless, some of the chemical abundances are similar between P1836849 and SDSS J102915, e.g., $\alpha$-elements (other than Ti).
The same is not true when the stellar chemistries are compared with 2MASS J18082002 and the four SkyMapper stars, which has very different abundances for Na, Cr, Mn, and possibly Al, Sc, and Ti.

Were these stars born in the same formation site?  It seems unlikely, despite some chemical and/or dynamical similarities discussed above.  
Furthermore, if they have been orbiting the MW since the early Galactic assembly, we can expect that they would have experienced many perturbations over cosmic time (\eg Gaia-Sausage-Enceladus, \citealt{Belokurov18,Helmi18}), which could have heated or altered their orbital configurations \citep[\eg][]{Navarro2018, DiMatteo19}.  
Their orbits may also have been affected by secular and non-linear interactions between the rotating MW bar and its spiral arms
\citep{Minchev10,Sestito20}. 
An investigation into a common origin for EMP stars on prograde quasi-circular orbits in the Galactic plane will require larger statistical samples than presented here.

\begin{figure*}
\includegraphics[width=1\textwidth]{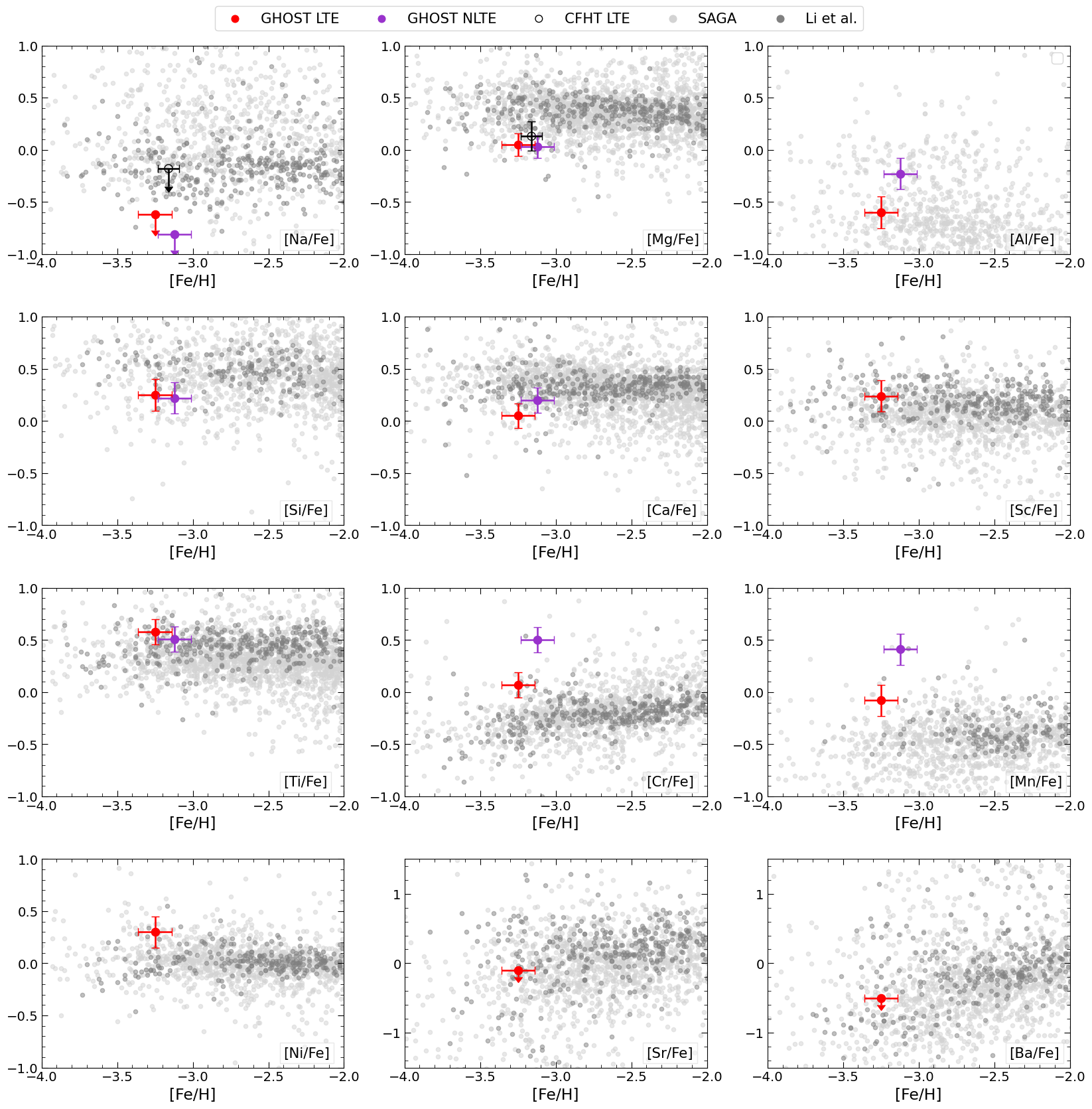}
\caption{
Chemical abundances of P1836849 compared with metal-poor stars in the MW (halo) from the homogeneous high-resolution spectroscopic study by 
\citet[][dark gray markers]{LiH21} and other stars in the SAGA database \citep[][and references within]{Suda08} (light gray markers). Red and purple symbols represent our LTE and NLTE-corrected abundances for P1836849, respectively. 
}
\label{Fig:chems}
\end{figure*}

\subsection{Comparisons with other EMP stars in the MW halo, Sculptor, and UFD galaxies }
\label{sect:sclufd}

The chemical abundances of P1836849 are compared to a compilation of stars of similar metallicity in the MW halo in Fig.~\ref{Fig:chems}.  
This includes chemical abundances of stars gathered from the literature in the Stellar Abundances for Galactic Archaeology database\footnote{\url{http://sagadatabase.jp}} \citep[SAGA, light grey circles;][]{Suda08}, and the high-resolution spectroscopic dataset taken with HDS at the Subaru Telescope and analysed homogeneously by \citet[][dark grey circles]{Li22}.
It is clear that the chemistry of P1836849 does not resemble the majority of EMP stars in the MW halo.   For example;

\begin{itemize}
    \item{ The [Na/Fe] upper limits found for P1836840 are extremely low compared to nearly all MW halo stars, in both LTE and NLTE.}
    \item{ The $\alpha$-elements (Mg, Si, Ca) are only consistent with the lowest values found in the MW halo stars.}
    \item { The Cr and Mn abundances are higher than the majority of the MW halo stars, and {\it very high} after NLTE corrections are applied. }
\end{itemize}

\noindent These are unlikely due to systematic errors in the NLTE corrections, as many of the halo stars are nearby F and G dwarfs with small to negligible NLTE corrections for most of their spectral lines.

\begin{figure*}
\includegraphics[width=1\textwidth]{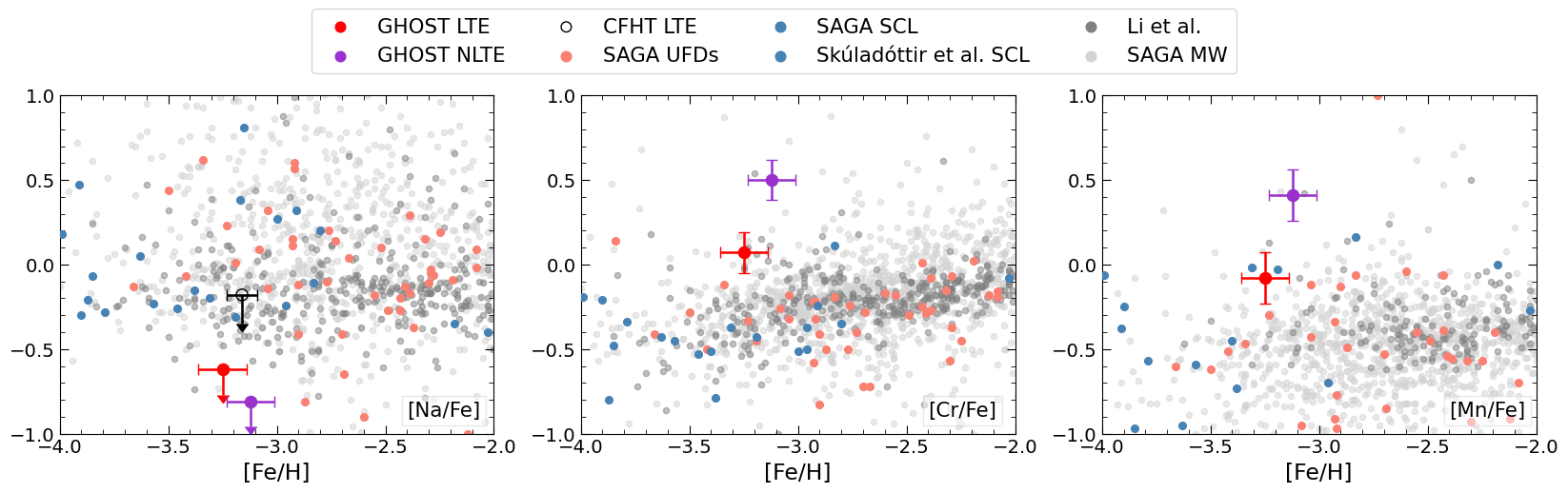}
\caption{A comparison of [Na/Fe], [Cr/Fe], and [Mn/Fe] of P1836849 to stars in the MW (symbols and sources same as in Fig.~\ref{Fig:chems}), to stars in the UFD galaxies as summarized in the SAGA database \citep[salmon markers; see text]{Suda2017}, and to stars in the classical dwarf galaxy Sculptor from both the SAGA database and \citet[][]{Skuladottir2023} (steelblue markers; see text).  
We note that the majority of the abundance ratios shown here have not been corrected for NLTE effects.
}
\label{Fig:ufds}
\end{figure*}

The chemistry of P1836849 can also be compared to EMP stars in nearby dwarf galaxies.  As an example, in Fig.~\ref{Fig:ufds} we compare our results to a sample of homogeneously analysed EMP stars in the classical ('textbook') dwarf galaxy Sculptor \citep[e.g., ][]{Hill2019, Skuladottir2023}.    
The $\alpha$-elements in Sculptor are slightly lower than EMP stars in the MW halo, which is typical of dwarf galaxies and has been discussed in terms of the slower star formation history of low mass satellites \citep{Venn04, Venn12, Tolstoy09, Jablonka15, Hill2019}.   Thus, the [$\alpha$/Fe] ratios in P1836849 are more similar to the EMP stars in Sculptor (not shown); however, P1836849 still stands out in Na and the iron-peak elements, as shown in Fig.~\ref{Fig:ufds}.  Note that we have included additional stars in Sculptor from the SAGA database \citep{Suda2017}, however most of those have [Fe/H]$>-2.5$.  

Our results for P1836849 are also compared to EMP stars in the UFD galaxies collected in the SAGA database by \citet[][and references therein]{Suda2017}.  This includes EMP stars in BooI, BooII, CVnI, CVnII, Com, GruI, Her, Hor, LeoII, LeoIV, LeoT, PscII, RetII, Seg1, Seg2, TriII, TucII, TucIII, UMaII, \& WilI.  As seen in Fig.~\ref{Fig:ufds},  again, the only elements that stand out in P1836849 are Na, Cr, and Mn when compared to the EMP stars in the UFDs. 

Thus, in general;
\begin{itemize}
    \item{ The [Na/Fe] upper limits found for P1836840 are much lower than the EMP stars in Sculptor and the majority of EMP stars in the UFDs. }
    \item { The LTE and NLTE Cr abundances are higher than for the stars in Sculptor ($\gtrsim$0.5 dex) and in UFDs ($\gtrsim$0.2 dex).}
    \item { The Mn (and possibly Ni, not shown) abundances are higher than the majority of the comparison stars in Sculptor and the UFDs.  A homogeneous analysis of \ion{Mn}{I} with NLTE corrections may be necessary to further compare these stellar populations. }
\end{itemize}

Finally, we note that \cite{Skuladottir21, Skuladottir2023} suggest that one EMP star AS0039 in Sculptor has a chemical abundance pattern that resembles enrichment from theoretical yields of a zero-metallicity hypernova progenitor (of mass M = 20M$_\odot$), solidifying this galaxy as a benchmark for understanding the first supernovae in the Universe.  In the next section, we compare P1836849 to theoretical yields from Population III supernovae.


\subsection{StarFit result }

To examine if the chemical abundance ratios in P1836849 could be reproduced by the predicted nucleosynthetic yields from Population III supernovae (SNe) and hypernovae (HNe), our LTE and NLTE abundances are compared to theoretical yields from \citet{Heger10} and \citet{Heger12} using 
the web version of  \textsc{StarFit}\footnote{\url{https://starfit.org/}} (v0.19.1).
These models predict the nucleosynthetic products of massive metal-free stars, without mass loss or  rotation, and with a range of explosion energies and mixing fractions.  
The fallback models (S4) used in this work have masses from 10 to 100 M$_\odot$, explosion energies ranging from  0.3 x 10$^{\rm 51}$ erg to 10$^{\rm 52}$ erg, and a range of mixing 
prescriptions.
\textsc{StarFit} can be used to search for a single SN or HN progenitor or a combination of SNe and HNe, providing a $\chi^{\rm 2}$ for the best fit to the observed abundances.  
This algorithm has been applied successfully to EMP stars in the literature \citep[e.g.,][]{Placco2016, Placco2020, Nordlander17, Skuladottir21}

\begin{figure}
\centering
\includegraphics[width=0.47\textwidth]
{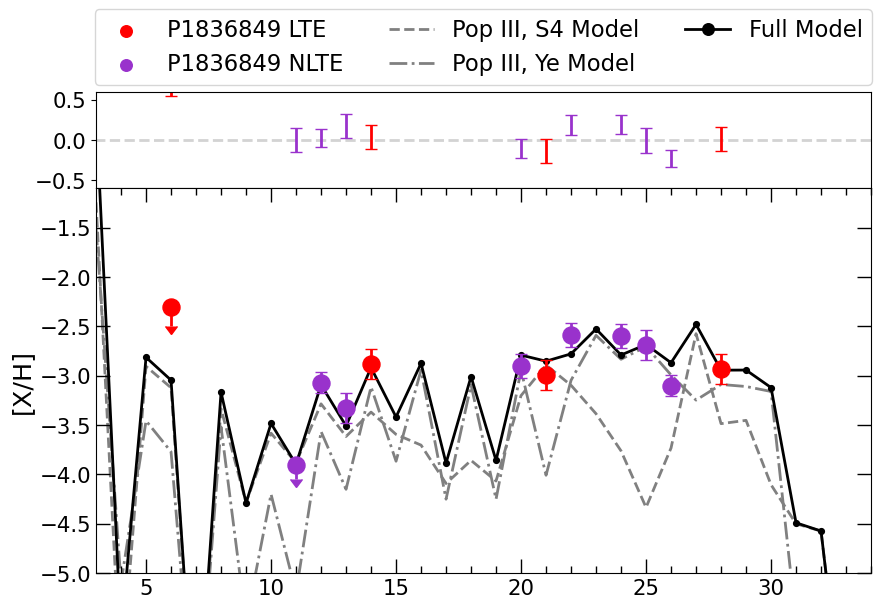}
\caption{P1836849 LTE (red) and NLTE (purple) chemical abundances vs atomic number compared to our best StarFit model (see text), which includes a low mass Pop III supernova (10 M$_\odot$, S4 models) and a low mass Pop III hypernova (17 M$_\odot$, Ye models).  NLTE abundances were used whenever possible, and LTE abundances are shown for the remaining elements, with  1$\sigma$ residuals from the model shown across the top.
}
\label{Fig:starfit}
\end{figure}

At first, the \textsc{StarFit} solutions to the chemistry of P1836849 appeared to be poorly constrained, due to  insufficient chemical data, especially for the neutron-capture elements. \textsc{StarFit} either struggled to converge, produced a range of models with satisfactory fits, or failed to converge to the same solution after repeated trials with the same input parameters. 
To improve the application of \textsc{StarFit}, we reduced the search parameters to only 1-3 SNe and/or HNe from the updated fallback models by \citet{Heger12}, and only fit the data from H to Ni
using the \textsc{Genetic Algorithm} and a 60 second time limit. 
Our best fit to the NLTE abundances for P1836849 is shown in Fig.~\ref{Fig:starfit}, which includes a 10.2 M$_\odot$ SN model with explosion energy B = 1.8 x 10$^{51}$ erg and mixing parameter log($f_{\rm mix}$)$= -1.4$ from the S4 models combined with a 17.1 M$_\odot$ HN model with higher explosion energy B = 10.0 x 10$^{51}$ erg and the same mixing parameter from the Ye models. This fit provides a $\chi^2$=0.94, compared to either model (or other single models) independently, where $\chi^2>2$. 
We note that the high abundances of Cr and Mn in our results are produced by the HN event, i.e., from incomplete Si-burning layers.

This result is consistent with the analyses by \cite{Skuladottir21, Skuladottir2023} for EMP stars in Sculptor (see Section~\ref{sect:sclufd}), and also \cite{Ishigaki18, Ishigaki14} who found
that the abundance patterns of $\sim$200 EMP stars in the MW halo are best-fit by SN with mass $<$40 M$_\odot$ and/or HN with mass =25 M$_\odot$.  This result led them to suggest that the masses of the first stars responsible for the early metal enrichment in the Galaxy were not extremely high, either because high-mass first stars were rare, they directly collapsed into a black hole without ejecting heavy elements, or supernova explosions from higher-mass Population III stars may have inhibited their formation. Studies of EMP stars in other nearby galaxies, and \textit{old} EMP stars in the MW, can address these options, i.e., where kinematic information in target selection may help in the future.

\section{Conclusions}\label{sec:conclusions}

As part of the commissioning of the new Gemini
High-resolution Optical SpecTrograph (GHOST), we have observed an EMP star with a prograde quasi-circular orbit in the Galactic plane, Pristine\_183.6849+04.8619 (P1836849), during the Science Verification stage. The exquisite throughput of GHOST has enabled a detailed spectral analysis of features from 3700 - 11000 \AA\ of many chemical elements (Mg,
Al, Si, Ca, Sc, Ti, Cr, Mn, Fe, Ni), and has provided valuable upper limits for others (Na, Sr, Ba).  This star is extremely metal-poor ([Fe/H]=-3.3$\pm$0.1) compared to other stars with MW planar orbits, and shows unusually low [Na/Fe] and high [Cr/Fe] and [Mn/Fe] compared with other EMP stars in the MW halo, Sculptor, and UFD galaxies. 
A simple comparison of our NLTE abundances to theoretical yields from supernova models suggests that only two low mass Population III objects are needed to reproduce the abundance pattern: one 10 M$_\odot$ supernova and one 17 M$_\odot$ hypernova (reduced $\chi^2<1$). 
Our analysis of P1836849 contributes to the growing evidence that the earliest stages of chemical enrichment in the Universe were dominated by low mass Population III supernovae and hypernovae.

\section*{Acknowledgements}
We acknowledge and respect the l\textschwa\textvbaraccent {k}$^{\rm w}$\textschwa\ng{}\textschwa n peoples on whose traditional territory the University of Victoria stands and the Songhees, Esquimalt and $\ubar{\rm W}$S\'ANE\'C  peoples whose historical relationships with the land continue to this day.

AD, KAV and FS thank the National Sciences and Engineering Research Council of Canada for funding through the Discovery Grants and USRA programs. 
FS also thanks the Dr. Margaret "Marmie" Perkins Hess postdoctoral fellowship for funding his work at the University of Victoria. 
NFM gratefully acknowledge support from the French National Research Agency (ANR) funded project ``Pristine'' (ANR-18-CE31-0017) along with funding from the European Research Council (ERC) under the European Unions Horizon 2020 research and innovation programme (grant agreement No. 834148).
EM acknowledges funding from FAPEMIG under project number APQ-02493-22 and a research productivity grant number 309829/2022-4 awarded by the CNPq, Brazil.
We would like to thank the anonymous referee who provided several helpful comments and suggestions that improved this paper.

Based on observations obtained under Program ID GS-2023A-SV-19, at the International Gemini Observatory, a program of NSF’s NOIRLab, which is managed by the Association of Universities for Research in Astronomy (AURA) under a cooperative agreement with the National Science Foundation on behalf of the Gemini Observatory partnership: the National Science Foundation (United States), National Research Council (Canada), Agencia Nacional de Investigaci\'{o}n y Desarrollo (Chile), Ministerio de Ciencia, Tecnolog\'{i}a e Innovaci\'{o}n (Argentina), Minist\'{e}rio da Ci\^{e}ncia, Tecnologia, Inova\c{c}\~{o}es e Comunica\c{c}\~{o}es (Brazil), and Korea Astronomy and Space Science Institute (Republic of Korea).

GHOST was built by a collaboration between Australian Astronomical Optics at Macquarie University, National Research Council Herzberg of Canada, and the Australian National University, and funded by the International Gemini partnership. The instrument scientist is Dr.\,Alan W.\,McConnachie at NRC, and the instrument team is also led by Dr.\,J.\,Gordon Robertson (at AAO), and Dr.\,Michael Ireland (at ANU). 
The authors would like to acknowledge the contributions of the GHOST instrument build team, the Gemini GHOST instrument team, the full SV team, and the rest of the Gemini operations team that were involved in making the SV observations a success.

This work has made use of data from the European Space Agency (ESA) mission \textit{Gaia} (\url{https://www.cosmos.esa.int/gaia}), processed by the \textit{Gaia} Data Processing and Analysis Consortium (DPAC, \url{https://www.cosmos.esa.int/web/gaia/dpac/consortium}). Funding for the DPAC has been provided by national institutions, in particular the institutions participating in the \textit{Gaia} Multilateral Agreement.

Funding for the Sloan Digital Sky 
Survey IV has been provided by the 
Alfred P. Sloan Foundation, the U.S. 
Department of Energy Office of 
Science, and the Participating 
Institutions. 
SDSS-IV acknowledges support and 
resources from the Center for High 
Performance Computing  at the 
University of Utah. The SDSS 
website is www.sdss4.org.
SDSS-IV is managed by the 
Astrophysical Research Consortium 
for the Participating Institutions 
of the SDSS Collaboration including 
the Brazilian Participation Group, 
the Carnegie Institution for Science, 
Carnegie Mellon University, Center for 
Astrophysics | Harvard \& 
Smithsonian, the Chilean Participation 
Group, the French Participation Group, 
Instituto de Astrof\'isica de 
Canarias, The Johns Hopkins 
University, Kavli Institute for the 
Physics and Mathematics of the 
Universe (IPMU) / University of 
Tokyo, the Korean Participation Group, 
Lawrence Berkeley National Laboratory, 
Leibniz Institut f\"ur Astrophysik 
Potsdam (AIP),  Max-Planck-Institut 
f\"ur Astronomie (MPIA Heidelberg), 
Max-Planck-Institut f\"ur 
Astrophysik (MPA Garching), 
Max-Planck-Institut f\"ur 
Extraterrestrische Physik (MPE), 
National Astronomical Observatories of 
China, New Mexico State University, 
New York University, University of 
Notre Dame, Observat\'ario 
Nacional / MCTI, The Ohio State 
University, Pennsylvania State 
University, Shanghai 
Astronomical Observatory, United 
Kingdom Participation Group, 
Universidad Nacional Aut\'onoma 
de M\'exico, University of Arizona, 
University of Colorado Boulder, 
University of Oxford, University of 
Portsmouth, University of Utah, 
University of Virginia, University 
of Washington, University of 
Wisconsin, Vanderbilt University, 
and Yale University.

This work made extensive use of \textsc{TOPCAT} \citep{Taylor05}. This research has made use of the \textsc{SIMBAD} database, operated at CDS, Strasbourg, France \citep{Wenger00}.

\section*{Data Availability}
GHOST Science Verification spectra are available at the Gemini Archive web page \url{https://archive.gemini.edu/searchform}. All data are incorporated into the article.

\bibliographystyle{mn2e}
\bibliography{ms}

\begin{thebibliography}{127}
\expandafter\ifx\csname natexlab\endcsname\relax\def\natexlab#1{#1}\fi

\bibitem[{{Abadi} {et~al}\mbox{.}(2003){Abadi}, {Navarro}, {Steinmetz}, \& {Eke}}]{Abadi03}
{Abadi} M.~G., {Navarro} J.~F., {Steinmetz} M., {Eke} V.~R., 2003, \apj, 597, 21

\bibitem[{{Abdurro'uf} {et~al}\mbox{.}(2022{\natexlab{a}}){Abdurro'uf}, {Accetta}, {Aerts}, {Silva Aguirre}, {Ahumada}, {Ajgaonkar}, {Filiz Ak}, {Alam}, {Allende Prieto}, {Almeida}, {Anders}, {Anderson}, {Andrews}, {Anguiano}, {Aquino-Ort{\'\i}z}, {Arag{\'o}n-Salamanca}, {Argudo-Fern{\'a}ndez}, {Ata}, {Aubert}, {Avila-Reese}, {Badenes}, {Barb{\'a}}, {Barger}, {Barrera-Ballesteros}, {Beaton}, {Beers}, {Belfiore}, {Bender}, {Bernardi}, {Bershady}, {Beutler}, {Bidin}, {Bird}, {Bizyaev}, {Blanc}, {Blanton}, {Boardman}, {Bolton}, {Boquien}, {Borissova}, {Bovy}, {Brandt}, {Brown}, {Brownstein}, {Brusa}, {Buchner}, {Bundy}, {Burchett}, {Bureau}, {Burgasser}, {Cabang}, {Campbell}, {Cappellari}, {Carlberg}, {Wanderley}, {Carrera}, {Cash}, {Chen}, {Chen}, {Cherinka}, {Chiappini}, {Choi}, {Chojnowski}, {Chung}, {Clerc}, {Cohen}, {Comerford}, {Comparat}, {da Costa}, {Covey}, {Crane}, {Cruz-Gonzalez}, {Culhane}, {Cunha}, {Dai}, {Damke}, {Darling}, {Davidson}, {Davies}, {Dawson}, {De Lee}, {Diamond-Stanic},
  {Cano-D{\'\i}az}, {S{\'a}nchez}, {Donor}, {Duckworth}, {Dwelly}, {Eisenstein}, {Elsworth}, {Emsellem}, {Eracleous}, {Escoffier}, {Fan}, {Farr}, {Feng}, {Fern{\'a}ndez-Trincado}, {Feuillet}, {Filipp}, {Fillingham}, {Frinchaboy}, {Fromenteau}, {Galbany}, {Garc{\'\i}a}, {Garc{\'\i}a-Hern{\'a}ndez}, {Ge}, {Geisler}, {Gelfand}, {G{\'e}ron}, {Gibson}, {Goddy}, {Godoy-Rivera}, {Grabowski}, {Green}, {Greener}, {Grier}, {Griffith}, {Guo}, {Guy}, {Hadjara}, {Harding}, {Hasselquist}, {Hayes}, {Hearty}, {Hern{\'a}ndez}, {Hill}, {Hogg}, {Holtzman}, {Horta}, {Hsieh}, {Hsu}, {Hsu}, {Huber}, {Huertas-Company}, {Hutchinson}, {Hwang}, {Ibarra-Medel}, {Chitham}, {Ilha}, {Imig}, {Jaekle}, {Jayasinghe}, {Ji}, {Johnson}, {Jones}, {J{\"o}nsson}, {Katkov}, {Khalatyan}, {Kinemuchi}, {Kisku}, {Knapen}, {Kneib}, {Kollmeier}, {Kong}, {Kounkel}, {Kreckel}, {Krishnarao}, {Lacerna}, {Lane}, {Langgin}, {Lavender}, {Law}, {Lazarz}, {Leung}, {Leung}, {Lewis}, {Li}, {Li}, {Lian}, {Liang}, {Lin}, {Lin}, {Lin}, {Lintott}, {Long},
  {Longa-Pe{\~n}a}, {L{\'o}pez-Cob{\'a}}, {Lu}, {Lundgren}, {Luo}, {Mackereth}, {de la Macorra}, {Mahadevan}, {Majewski}, {Manchado}, {Mandeville}, {Maraston}, {Margalef-Bentabol}, {Masseron}, {Masters}, {Mathur}, {McDermid}, {Mckay}, {Merloni}, {Merrifield}, {Meszaros}, {Miglio}, {Di Mille}, {Minniti}, {Minsley}, {Monachesi}, {Moon}, {Mosser}, {Mulchaey}, {Muna}, {Mu{\~n}oz}, {Myers}, {Myers}, {Nadathur}, {Nair}, {Nandra}, {Neumann}, {Newman}, {Nidever}, {Nikakhtar}, {Nitschelm}, {O'Connell}, {Garma-Oehmichen}, {Luan Souza de Oliveira}, {Olney}, {Oravetz}, {Ortigoza-Urdaneta}, {Osorio}, {Otter}, {Pace}, {Padilla}, {Pan}, {Pan}, {Parikh}, {Parker}, {Peirani}, {Pe{\~n}a Ram{\'\i}rez}, {Penny}, {Percival}, {Perez-Fournon}, {Pinsonneault}, {Poidevin}, {Poovelil}, {Price-Whelan}, {B{\'a}rbara de Andrade Queiroz}, {Raddick}, {Ray}, {Rembold}, {Riddle}, {Riffel}, {Riffel}, {Rix}, {Robin}, {Rodr{\'\i}guez-Puebla}, {Roman-Lopes}, {Rom{\'a}n-Z{\'u}{\~n}iga}, {Rose}, {Ross}, {Rossi}, {Rubin}, {Salvato}, {S{\'a}nchez},
  {S{\'a}nchez-Gallego}, {Sanderson}, {Santana Rojas}, {Sarceno}, {Sarmiento}, {Sayres}, {Sazonova}, {Schaefer}, {Schiavon}, {Schlegel}, {Schneider}, {Schultheis}, {Schwope}, {Serenelli}, {Serna}, {Shao}, {Shapiro}, {Sharma}, {Shen}, {Shetrone}, {Shu}, {Simon}, {Skrutskie}, {Smethurst}, {Smith}, {Sobeck}, {Spoo}, {Sprague}, {Stark}, {Stassun}, {Steinmetz}, {Stello}, {Stone-Martinez}, {Storchi-Bergmann}, {Stringfellow}, {Stutz}, {Su}, {Taghizadeh-Popp}, {Talbot}, {Tayar}, {Telles}, {Teske}, {Thakar}, {Theissen}, {Tkachenko}, {Thomas}, {Tojeiro}, {Hernandez Toledo}, {Troup}, {Trump}, {Trussler}, {Turner}, {Tuttle}, {Unda-Sanzana}, {V{\'a}zquez-Mata}, {Valentini}, {Valenzuela}, {Vargas-Gonz{\'a}lez}, {Vargas-Maga{\~n}a}, {Alfaro}, {Villanova}, {Vincenzo}, {Wake}, {Warfield}, {Washington}, {Weaver}, {Weijmans}, {Weinberg}, {Weiss}, {Westfall}, {Wild}, {Wilde}, {Wilson}, {Wilson}, {Wilson}, {Wolf}, {Wood-Vasey}, {Yan}, {Zamora}, {Zasowski}, {Zhang}, {Zhao}, {Zheng}, {Zheng}, \& {Zhu}}]{sdssdr17}
{Abdurro'uf} {et~al.}, 2022{\natexlab{a}}, \apjs, 259, 35

\bibitem[{{Abdurro'uf} {et~al}\mbox{.}(2022{\natexlab{b}}){Abdurro'uf}, {Accetta}, {Aerts}, {Silva Aguirre}, {Ahumada}, {Ajgaonkar}, {Filiz Ak}, {Alam}, {Allende Prieto}, {Almeida}, \& et~al.}]{APOGEEDR17}
---, 2022{\natexlab{b}}, \apjs, 259, 35

\bibitem[{{Aguado} {et~al}\mbox{.}(2019){Aguado}, {Youakim}, {Gonz{\'a}lez Hern{\'a}ndez}, {Allende Prieto}, {Starkenburg}, {Martin}, {Bonifacio}, {Arentsen}, {Caffau}, {Peralta de Arriba}, {Sestito}, {Garcia-Dias}, {Fantin}, {Hill}, {Jablonca}, {Jahandar}, {Kielty}, {Longeard}, {Lucchesi}, {S{\'a}nchez-Janssen}, {Osorio}, {Palicio}, {Tolstoy}, {Wilson}, {C{\^o}t{\'e}}, {Kordopatis}, {Lardo}, {Navarro}, {Thomas}, \& {Venn}}]{Aguado19}
{Aguado} D.~S. {et~al.}, 2019, \mnras, 490, 2241

\bibitem[{{Aoki} {et~al}\mbox{.}(2013){Aoki}, {Beers}, {Lee}, {Honda}, {Ito}, {Takada-Hidai}, {Frebel}, {Suda}, {Fujimoto}, {Carollo}, \& {Sivarani}}]{Aoki13}
{Aoki} W. {et~al.}, 2013, \aj, 145, 13

\bibitem[{{Asplund} {et~al}\mbox{.}(2009){Asplund}, {Grevesse}, {Sauval}, \& {Scott}}]{Asplund09}
{Asplund} M., {Grevesse} N., {Sauval} A.~J., {Scott} P., 2009, \araa, 47, 481

\bibitem[{{Battistini} \& {Bensby}(2015)}]{Battistini2015}
{Battistini} C., {Bensby} T., 2015, \aap, 577, A9

\bibitem[{{Battistini} \& {Bensby}(2016)}]{Battistini2016}
---, 2016, \aap, 586, A49

\bibitem[{{Beaton} {et~al}\mbox{.}(2021){Beaton}, {Oelkers}, {Hayes}, {Covey}, {Chojnowski}, {De Lee}, {Sobeck}, {Majewski}, {Cohen}, {Fern{\'a}ndez-Trincado}, {Longa-Pe{\~n}a}, {O'Connell}, {Santana}, {Stringfellow}, {Zasowski}, {Aerts}, {Anguiano}, {Bender}, {Ca{\~n}as}, {Cunha}, {Donor}, {Fleming}, {Frinchaboy}, {Feuillet}, {Harding}, {Hasselquist}, {Holtzman}, {Johnson}, {Kollmeier}, {Kounkel}, {Mahadevan}, {Price-Whelan}, {Rojas-Arriagada}, {Rom{\'a}n-Z{\'u}{\~n}iga}, {Schlafly}, {Schultheis}, {Shetrone}, {Simon}, {Stassun}, {Stutz}, {Tayar}, {Teske}, {Tkachenko}, {Troup}, {Albareti}, {Bizyaev}, {Bovy}, {Burgasser}, {Comparat}, {Downes}, {Geisler}, {Inno}, {Manchado}, {Ness}, {Pinsonneault}, {Prada}, {Roman-Lopes}, {Simonian}, {Smith}, {Yan}, \& {Zamora}}]{Beaton2021}
{Beaton} R.~L. {et~al.}, 2021, \aj, 162, 302

\bibitem[{{Belokurov} {et~al}\mbox{.}(2018){Belokurov}, {Erkal}, {Evans}, {Koposov}, \& {Deason}}]{Belokurov18}
{Belokurov} V., {Erkal} D., {Evans} N.~W., {Koposov} S.~E., {Deason} A.~J., 2018, \mnras, 478, 611

\bibitem[{{Belokurov} \& {Kravtsov}(2022)}]{Belokurov22}
{Belokurov} V., {Kravtsov} A., 2022, \mnras, 514, 689

\bibitem[{{Belokurov} \& {Kravtsov}(2023)}]{Belokurov23}
---, 2023, arXiv e-prints, arXiv:2306.00060

\bibitem[{{Bensby}, {Feltzing} \& {Oey}(2014){Bensby}, {Feltzing}, \& {Oey}}]{Bensby2014}
{Bensby} T., {Feltzing} S., {Oey} M.~S., 2014, \aap, 562, A71

\bibitem[{{Bergemann}(2011)}]{Bergemann2011}
{Bergemann} M., 2011, \mnras, 413, 2184

\bibitem[{{Bergemann} \& {Cescutti}(2010)}]{Bergemann2010b}
{Bergemann} M., {Cescutti} G., 2010, \aap, 522, A9

\bibitem[{{Bergemann} {et~al}\mbox{.}(2017){Bergemann}, {Collet}, {Amarsi}, {Kovalev}, {Ruchti}, \& {Magic}}]{Bergemann2017}
{Bergemann} M., {Collet} R., {Amarsi} A.~M., {Kovalev} M., {Ruchti} G., {Magic} Z., 2017, \apj, 847, 15

\bibitem[{{Bergemann} {et~al}\mbox{.}(2019){Bergemann}, {Gallagher}, {Eitner}, {Bautista}, {Collet}, {Yakovleva}, {Mayriedl}, {Plez}, {Carlsson}, {Leenaarts}, {Belyaev}, \& {Hansen}}]{Bergemann2019}
{Bergemann} M. {et~al.}, 2019, \aap, 631, A80

\bibitem[{{Bergemann} {et~al}\mbox{.}(2013){Bergemann}, {Kudritzki}, {W{\"u}rl}, {Plez}, {Davies}, \& {Gazak}}]{Bergemann2013}
{Bergemann} M., {Kudritzki} R.-P., {W{\"u}rl} M., {Plez} B., {Davies} B., {Gazak} Z., 2013, \apj, 764, 115

\bibitem[{{Bland-Hawthorn} \& {Gerhard}(2016)}]{BlandHawthorn16}
{Bland-Hawthorn} J., {Gerhard} O., 2016, \araa, 54, 529

\bibitem[{{Bovy}(2015)}]{Bovy15}
{Bovy} J., 2015, \apjs, 216, 29

\bibitem[{Bowen \& Vaughan(1973)}]{Bowen1973}
Bowen I.~S., Vaughan A.~H., 1973, Appl. Opt., 12, 1430

\bibitem[{{Bullock} \& {Johnston}(2005)}]{Bullock2005}
{Bullock} J.~S., {Johnston} K.~V., 2005, \apj, 635, 931

\bibitem[{{Caffau} {et~al}\mbox{.}(2012){Caffau}, {Bonifacio}, {Fran{\c c}ois}, {Spite}, {Spite}, {Zaggia}, {Ludwig}, {Steffen}, {Mashonkina}, {Monaco}, {Sbordone}, {Molaro}, {Cayrel}, {Plez}, {Hill}, {Hammer}, \& {Randich}}]{Caffau12}
{Caffau} E. {et~al.}, 2012, \aap, 542, A51

\bibitem[{{Caffau} {et~al}\mbox{.}(2011){Caffau}, {Bonifacio}, {Fran{\c{c}}ois}, {Sbordone}, {Monaco}, {Spite}, {Spite}, {Ludwig}, {Cayrel}, {Zaggia}, {Hammer}, {Randich}, {Molaro}, \& {Hill}}]{Caffau11}
---, 2011, \nat, 477, 67

\bibitem[{{Choi} {et~al}\mbox{.}(2016){Choi}, {Dotter}, {Conroy}, {Cantiello}, {Paxton}, \& {Johnson}}]{Choi16}
{Choi} J., {Dotter} A., {Conroy} C., {Cantiello} M., {Paxton} B., {Johnson} B.~D., 2016, \apj, 823, 102

\bibitem[{{Cordoni} {et~al}\mbox{.}(2021){Cordoni}, {Da Costa}, {Yong}, {Mackey}, {Marino}, {Monty}, {Nordlander}, {Norris}, {Asplund}, {Bessell}, {Casey}, {Frebel}, {Lind}, {Murphy}, {Schmidt}, {Gao}, {Xylakis-Dornbusch}, {Amarsi}, \& {Milone}}]{Cordoni21}
{Cordoni} G. {et~al.}, 2021, \mnras, 503, 2539

\bibitem[{{Das}, {Hawkins} \& {Jofr{\'e}}(2020){Das}, {Hawkins}, \& {Jofr{\'e}}}]{Das20}
{Das} P., {Hawkins} K., {Jofr{\'e}} P., 2020, \mnras, 493, 5195

\bibitem[{{Di Matteo} {et~al}\mbox{.}(2019){Di Matteo}, {Haywood}, {Lehnert}, {Katz}, {Khoperskov}, {Snaith}, {G{\'o}mez}, \& {Robichon}}]{DiMatteo19}
{Di Matteo} P., {Haywood} M., {Lehnert} M.~D., {Katz} D., {Khoperskov} S., {Snaith} O.~N., {G{\'o}mez} A., {Robichon} N., 2019, \aap, 632, A4

\bibitem[{{Donati} {et~al}\mbox{.}(2006){Donati}, {Catala}, {Landstreet}, \& {Petit}}]{Donati2006}
{Donati} J.~F., {Catala} C., {Landstreet} J.~D., {Petit} P., 2006, in Astronomical Society of the Pacific Conference Series, Vol. 358, Solar Polarization 4, {Casini} R., {Lites} B.~W., eds., p. 362

\bibitem[{{Dotter}(2016)}]{Dotter16}
{Dotter} A., 2016, \apjs, 222, 8

\bibitem[{{El-Badry} {et~al}\mbox{.}(2018){El-Badry}, {Bland-Hawthorn}, {Wetzel}, {Quataert}, {Weisz}, {Boylan-Kolchin}, {Hopkins}, {Faucher-Gigu{\`e}re}, {Kere{\v s}}, \& {Garrison-Kimmel}}]{ElBadry18}
{El-Badry} K. {et~al.}, 2018, \mnras, 480, 652

\bibitem[{{Fern{\'a}ndez-Alvar} {et~al}\mbox{.}(2021){Fern{\'a}ndez-Alvar}, {Kordopatis}, {Hill}, {Starkenburg}, {Viswanathan}, {Martin}, {Thomas}, {Navarro}, {Malhan}, {Sestito}, {Gonz{\'a}lez Hern{\'a}ndez}, \& {Carlberg}}]{FernandezAlvar2021}
{Fern{\'a}ndez-Alvar} E. {et~al.}, 2021, \mnras, 508, 1509

\bibitem[{{Frebel}, {Kirby} \& {Simon}(2010){Frebel}, {Kirby}, \& {Simon}}]{Frebel10}
{Frebel} A., {Kirby} E.~N., {Simon} J.~D., 2010, \nat, 464, 72

\bibitem[{{Gaia Collaboration} {et~al}\mbox{.}(2018){Gaia Collaboration}, {Brown}, {Vallenari}, {Prusti}, {de Bruijne}, {Babusiaux}, {Bailer-Jones}, {Biermann}, {Evans}, {Eyer}, {Jansen}, {Jordi}, {Klioner}, {Lammers}, {Lindegren}, {Luri}, {Mignard}, {Panem}, {Pourbaix}, {Randich}, {Sartoretti}, {Siddiqui}, {Soubiran}, {van Leeuwen}, {Walton}, {Arenou}, {Bastian}, {Cropper}, {Drimmel}, {Katz}, {Lattanzi}, {Bakker}, {Cacciari}, {Casta{\~n}eda}, {Chaoul}, {Cheek}, {De Angeli}, {Fabricius}, {Guerra}, {Holl}, {Masana}, {Messineo}, {Mowlavi}, {Nienartowicz}, {Panuzzo}, {Portell}, {Riello}, {Seabroke}, {Tanga}, {Th{\'e}venin}, {Gracia-Abril}, {Comoretto}, {Garcia-Reinaldos}, {Teyssier}, {Altmann}, {Andrae}, {Audard}, {Bellas-Velidis}, {Benson}, {Berthier}, {Blomme}, {Burgess}, {Busso}, {Carry}, {Cellino}, {Clementini}, {Clotet}, {Creevey}, {Davidson}, {De Ridder}, {Delchambre}, {Dell'Oro}, {Ducourant}, {Fern{\'a}ndez-Hern{\'a}ndez}, {Fouesneau}, {Fr{\'e}mat}, {Galluccio}, {Garc{\'\i}a-Torres},
  {Gonz{\'a}lez-N{\'u}{\~n}ez}, {Gonz{\'a}lez-Vidal}, {Gosset}, {Guy}, {Halbwachs}, {Hambly}, {Harrison}, {Hern{\'a}ndez}, {Hestroffer}, {Hodgkin}, {Hutton}, {Jasniewicz}, {Jean-Antoine-Piccolo}, {Jordan}, {Korn}, {Krone-Martins}, {Lanzafame}, {Lebzelter}, {L{\"o}ffler}, {Manteiga}, {Marrese}, {Mart{\'\i}n-Fleitas}, {Moitinho}, {Mora}, {Muinonen}, {Osinde}, {Pancino}, {Pauwels}, {Petit}, {Recio-Blanco}, {Richards}, {Rimoldini}, {Robin}, {Sarro}, {Siopis}, {Smith}, {Sozzetti}, {S{\"u}veges}, {Torra}, {van Reeven}, {Abbas}, {Abreu Aramburu}, {Accart}, {Aerts}, {Altavilla}, {{\'A}lvarez}, {Alvarez}, {Alves}, {Anderson}, {Andrei}, {Anglada Varela}, {Antiche}, {Antoja}, {Arcay}, {Astraatmadja}, {Bach}, {Baker}, {Balaguer-N{\'u}{\~n}ez}, {Balm}, {Barache}, {Barata}, {Barbato}, {Barblan}, {Barklem}, {Barrado}, {Barros}, {Barstow}, {Bartholom{\'e} Mu{\~n}oz}, {Bassilana}, {Becciani}, {Bellazzini}, {Berihuete}, {Bertone}, {Bianchi}, {Bienaym{\'e}}, {Blanco-Cuaresma}, {Boch}, {Boeche}, {Bombrun}, {Borrachero},
  {Bossini}, {Bouquillon}, {Bourda}, {Bragaglia}, {Bramante}, {Breddels}, {Bressan}, {Brouillet}, {Br{\"u}semeister}, {Brugaletta}, {Bucciarelli}, {Burlacu}, {Busonero}, {Butkevich}, {Buzzi}, {Caffau}, {Cancelliere}, {Cannizzaro}, {Cantat-Gaudin}, {Carballo}, {Carlucci}, {Carrasco}, {Casamiquela}, {Castellani}, {Castro-Ginard}, {Charlot}, {Chemin}, {Chiavassa}, {Cocozza}, {Costigan}, {Cowell}, {Crifo}, {Crosta}, {Crowley}, {Cuypers}, {Dafonte}, {Damerdji}, {Dapergolas}, {David}, {David}, {de Laverny}, {De Luise}, {De March}, {de Martino}, {de Souza}, {de Torres}, {Debosscher}, {del Pozo}, {Delbo}, {Delgado}, {Delgado}, {Di Matteo}, {Diakite}, {Diener}, {Distefano}, {Dolding}, {Drazinos}, {Dur{\'a}n}, {Edvardsson}, {Enke}, {Eriksson}, {Esquej}, {Eynard Bontemps}, {Fabre}, {Fabrizio}, {Faigler}, {Falc{\~a}o}, {Farr{\`a}s Casas}, {Federici}, {Fedorets}, {Fernique}, {Figueras}, {Filippi}, {Findeisen}, {Fonti}, {Fraile}, {Fraser}, {Fr{\'e}zouls}, {Gai}, {Galleti}, {Garabato}, {Garc{\'\i}a-Sedano}, {Garofalo},
  {Garralda}, {Gavel}, {Gavras}, {Gerssen}, {Geyer}, {Giacobbe}, {Gilmore}, {Girona}, {Giuffrida}, {Glass}, {Gomes}, {Granvik}, {Gueguen}, {Guerrier}, {Guiraud}, {Guti{\'e}rrez-S{\'a}nchez}, {Haigron}, {Hatzidimitriou}, {Hauser}, {Haywood}, {Heiter}, {Helmi}, {Heu}, {Hilger}, {Hobbs}, {Hofmann}, {Holland}, {Huckle}, {Hypki}, {Icardi}, {Jan{\ss}en}, {Jevardat de Fombelle}, {Jonker}, {Juh{\'a}sz}, {Julbe}, {Karampelas}, {Kewley}, {Klar}, {Kochoska}, {Kohley}, {Kolenberg}, {Kontizas}, {Kontizas}, {Koposov}, {Kordopatis}, {Kostrzewa-Rutkowska}, {Koubsky}, {Lambert}, {Lanza}, {Lasne}, {Lavigne}, {Le Fustec}, {Le Poncin-Lafitte}, {Lebreton}, {Leccia}, {Leclerc}, {Lecoeur-Taibi}, {Lenhardt}, {Leroux}, {Liao}, {Licata}, {Lindstr{\o}m}, {Lister}, {Livanou}, {Lobel}, {L{\'o}pez}, {Managau}, {Mann}, {Mantelet}, {Marchal}, {Marchant}, {Marconi}, {Marinoni}, {Marschalk{\'o}}, {Marshall}, {Martino}, {Marton}, {Mary}, {Massari}, {Matijevi{\v{c}}}, {Mazeh}, {McMillan}, {Messina}, {Michalik}, {Millar}, {Molina}, {Molinaro},
  {Moln{\'a}r}, {Montegriffo}, {Mor}, {Morbidelli}, {Morel}, {Morris}, {Mulone}, {Muraveva}, {Musella}, {Nelemans}, {Nicastro}, {Noval}, {O'Mullane}, {Ord{\'e}novic}, {Ord{\'o}{\~n}ez-Blanco}, {Osborne}, {Pagani}, {Pagano}, {Pailler}, {Palacin}, {Palaversa}, {Panahi}, {Pawlak}, {Piersimoni}, {Pineau}, {Plachy}, {Plum}, {Poggio}, {Poujoulet}, {Pr{\v{s}}a}, {Pulone}, {Racero}, {Ragaini}, {Rambaux}, {Ramos-Lerate}, {Regibo}, {Reyl{\'e}}, {Riclet}, {Ripepi}, {Riva}, {Rivard}, {Rixon}, {Roegiers}, {Roelens}, {Romero-G{\'o}mez}, {Rowell}, {Royer}, {Ruiz-Dern}, {Sadowski}, {Sagrist{\`a} Sell{\'e}s}, {Sahlmann}, {Salgado}, {Salguero}, {Sanna}, {Santana-Ros}, {Sarasso}, {Savietto}, {Schultheis}, {Sciacca}, {Segol}, {Segovia}, {S{\'e}gransan}, {Shih}, {Siltala}, {Silva}, {Smart}, {Smith}, {Solano}, {Solitro}, {Sordo}, {Soria Nieto}, {Souchay}, {Spagna}, {Spoto}, {Stampa}, {Steele}, {Steidelm{\"u}ller}, {Stephenson}, {Stoev}, {Suess}, {Surdej}, {Szabados}, {Szegedi-Elek}, {Tapiador}, {Taris}, {Tauran}, {Taylor},
  {Teixeira}, {Terrett}, {Teyssandier}, {Thuillot}, {Titarenko}, {Torra Clotet}, {Turon}, {Ulla}, {Utrilla}, {Uzzi}, {Vaillant}, {Valentini}, {Valette}, {van Elteren}, {Van Hemelryck}, {van Leeuwen}, {Vaschetto}, {Vecchiato}, {Veljanoski}, {Viala}, {Vicente}, {Vogt}, {von Essen}, {Voss}, {Votruba}, {Voutsinas}, {Walmsley}, {Weiler}, {Wertz}, {Wevers}, {Wyrzykowski}, {Yoldas}, {{\v{Z}}erjal}, {Ziaeepour}, {Zorec}, {Zschocke}, {Zucker}, {Zurbach}, \& {Zwitter}}]{GaiaDR2}
{Gaia Collaboration} {et~al.}, 2018, \aap, 616, A1

\bibitem[{{Gaia Collaboration} {et~al}\mbox{.}(2021){Gaia Collaboration}, {Brown}, {Vallenari}, {Prusti}, {de Bruijne}, {Babusiaux}, {Biermann}, {Creevey}, {Evans}, {Eyer}, {Hutton}, {Jansen}, {Jordi}, {Klioner}, {Lammers}, {Lindegren}, {Luri}, {Mignard}, {Panem}, {Pourbaix}, {Randich}, {Sartoretti}, {Soubiran}, {Walton}, {Arenou}, {Bailer-Jones}, {Bastian}, {Cropper}, {Drimmel}, {Katz}, {Lattanzi}, {van Leeuwen}, {Bakker}, {Cacciari}, {Casta{\~n}eda}, {De Angeli}, {Ducourant}, {Fabricius}, {Fouesneau}, {Fr{\'e}mat}, {Guerra}, {Guerrier}, {Guiraud}, {Jean-Antoine Piccolo}, {Masana}, {Messineo}, {Mowlavi}, {Nicolas}, {Nienartowicz}, {Pailler}, {Panuzzo}, {Riclet}, {Roux}, {Seabroke}, {Sordo}, {Tanga}, {Th{\'e}venin}, {Gracia-Abril}, {Portell}, {Teyssier}, {Altmann}, {Andrae}, {Bellas-Velidis}, {Benson}, {Berthier}, {Blomme}, {Brugaletta}, {Burgess}, {Busso}, {Carry}, {Cellino}, {Cheek}, {Clementini}, {Damerdji}, {Davidson}, {Delchambre}, {Dell'Oro}, {Fern{\'a}ndez-Hern{\'a}ndez}, {Galluccio},
  {Garc{\'\i}a-Lario}, {Garcia-Reinaldos}, {Gonz{\'a}lez-N{\'u}{\~n}ez}, {Gosset}, {Haigron}, {Halbwachs}, {Hambly}, {Harrison}, {Hatzidimitriou}, {Heiter}, {Hern{\'a}ndez}, {Hestroffer}, {Hodgkin}, {Holl}, {Jan{\ss}en}, {Jevardat de Fombelle}, {Jordan}, {Krone-Martins}, {Lanzafame}, {L{\"o}ffler}, {Lorca}, {Manteiga}, {Marchal}, {Marrese}, {Moitinho}, {Mora}, {Muinonen}, {Osborne}, {Pancino}, {Pauwels}, {Petit}, {Recio-Blanco}, {Richards}, {Riello}, {Rimoldini}, {Robin}, {Roegiers}, {Rybizki}, {Sarro}, {Siopis}, {Smith}, {Sozzetti}, {Ulla}, {Utrilla}, {van Leeuwen}, {van Reeven}, {Abbas}, {Abreu Aramburu}, {Accart}, {Aerts}, {Aguado}, {Ajaj}, {Altavilla}, {{\'A}lvarez}, {{\'A}lvarez Cid-Fuentes}, {Alves}, {Anderson}, {Anglada Varela}, {Antoja}, {Audard}, {Baines}, {Baker}, {Balaguer-N{\'u}{\~n}ez}, {Balbinot}, {Balog}, {Barache}, {Barbato}, {Barros}, {Barstow}, {Bartolom{\'e}}, {Bassilana}, {Bauchet}, {Baudesson-Stella}, {Becciani}, {Bellazzini}, {Bernet}, {Bertone}, {Bianchi}, {Blanco-Cuaresma}, {Boch},
  {Bombrun}, {Bossini}, {Bouquillon}, {Bragaglia}, {Bramante}, {Breedt}, {Bressan}, {Brouillet}, {Bucciarelli}, {Burlacu}, {Busonero}, {Butkevich}, {Buzzi}, {Caffau}, {Cancelliere}, {C{\'a}novas}, {Cantat-Gaudin}, {Carballo}, {Carlucci}, {Carnerero}, {Carrasco}, {Casamiquela}, {Castellani}, {Castro-Ginard}, {Castro Sampol}, {Chaoul}, {Charlot}, {Chemin}, {Chiavassa}, {Cioni}, {Comoretto}, {Cooper}, {Cornez}, {Cowell}, {Crifo}, {Crosta}, {Crowley}, {Dafonte}, {Dapergolas}, {David}, {David}, {de Laverny}, {De Luise}, {De March}, {De Ridder}, {de Souza}, {de Teodoro}, {de Torres}, {del Peloso}, {del Pozo}, {Delbo}, {Delgado}, {Delgado}, {Delisle}, {Di Matteo}, {Diakite}, {Diener}, {Distefano}, {Dolding}, {Eappachen}, {Edvardsson}, {Enke}, {Esquej}, {Fabre}, {Fabrizio}, {Faigler}, {Fedorets}, {Fernique}, {Fienga}, {Figueras}, {Fouron}, {Fragkoudi}, {Fraile}, {Franke}, {Gai}, {Garabato}, {Garcia-Gutierrez}, {Garc{\'\i}a-Torres}, {Garofalo}, {Gavras}, {Gerlach}, {Geyer}, {Giacobbe}, {Gilmore}, {Girona},
  {Giuffrida}, {Gomel}, {Gomez}, {Gonzalez-Santamaria}, {Gonz{\'a}lez-Vidal}, {Granvik}, {Guti{\'e}rrez-S{\'a}nchez}, {Guy}, {Hauser}, {Haywood}, {Helmi}, {Hidalgo}, {Hilger}, {H{\l}adczuk}, {Hobbs}, {Holland}, {Huckle}, {Jasniewicz}, {Jonker}, {Juaristi Campillo}, {Julbe}, {Karbevska}, {Kervella}, {Khanna}, {Kochoska}, {Kontizas}, {Kordopatis}, {Korn}, {Kostrzewa-Rutkowska}, {Kruszy{\'n}ska}, {Lambert}, {Lanza}, {Lasne}, {Le Campion}, {Le Fustec}, {Lebreton}, {Lebzelter}, {Leccia}, {Leclerc}, {Lecoeur-Taibi}, {Liao}, {Licata}, {Lindstr{\o}m}, {Lister}, {Livanou}, {Lobel}, {Madrero Pardo}, {Managau}, {Mann}, {Marchant}, {Marconi}, {Marcos Santos}, {Marinoni}, {Marocco}, {Marshall}, {Martin Polo}, {Mart{\'\i}n-Fleitas}, {Masip}, {Massari}, {Mastrobuono-Battisti}, {Mazeh}, {McMillan}, {Messina}, {Michalik}, {Millar}, {Mints}, {Molina}, {Molinaro}, {Moln{\'a}r}, {Montegriffo}, {Mor}, {Morbidelli}, {Morel}, {Morris}, {Mulone}, {Munoz}, {Muraveva}, {Murphy}, {Musella}, {Noval}, {Ord{\'e}novic}, {Orr{\`u}},
  {Osinde}, {Pagani}, {Pagano}, {Palaversa}, {Palicio}, {Panahi}, {Pawlak}, {Pe{\~n}alosa Esteller}, {Penttil{\"a}}, {Piersimoni}, {Pineau}, {Plachy}, {Plum}, {Poggio}, {Poretti}, {Poujoulet}, {Pr{\v{s}}a}, {Pulone}, {Racero}, {Ragaini}, {Rainer}, {Raiteri}, {Rambaux}, {Ramos}, {Ramos-Lerate}, {Re Fiorentin}, {Regibo}, {Reyl{\'e}}, {Ripepi}, {Riva}, {Rixon}, {Robichon}, {Robin}, {Roelens}, {Rohrbasser}, {Romero-G{\'o}mez}, {Rowell}, {Royer}, {Rybicki}, {Sadowski}, {Sagrist{\`a} Sell{\'e}s}, {Sahlmann}, {Salgado}, {Salguero}, {Samaras}, {Sanchez Gimenez}, {Sanna}, {Santove{\~n}a}, {Sarasso}, {Schultheis}, {Sciacca}, {Segol}, {Segovia}, {S{\'e}gransan}, {Semeux}, {Shahaf}, {Siddiqui}, {Siebert}, {Siltala}, {Slezak}, {Smart}, {Solano}, {Solitro}, {Souami}, {Souchay}, {Spagna}, {Spoto}, {Steele}, {Steidelm{\"u}ller}, {Stephenson}, {S{\"u}veges}, {Szabados}, {Szegedi-Elek}, {Taris}, {Tauran}, {Taylor}, {Teixeira}, {Thuillot}, {Tonello}, {Torra}, {Torra}, {Turon}, {Unger}, {Vaillant}, {van Dillen}, {Vanel},
  {Vecchiato}, {Viala}, {Vicente}, {Voutsinas}, {Weiler}, {Wevers}, {Wyrzykowski}, {Yoldas}, {Yvard}, {Zhao}, {Zorec}, {Zucker}, {Zurbach}, \& {Zwitter}}]{gaiaedr3}
---, 2021, \aap, 649, A1

\bibitem[{{Gaia Collaboration} {et~al}\mbox{.}(2016){Gaia Collaboration}, {Prusti}, {de Bruijne}, {Brown}, {Vallenari}, {Babusiaux}, {Bailer-Jones}, {Bastian}, {Biermann}, {Evans}, \& et~al.}]{Gaia16}
---, 2016, \aap, 595, A1

\bibitem[{{Gaia Collaboration} {et~al}\mbox{.}(2023){Gaia Collaboration}, {Vallenari}, {Brown}, {Prusti}, {de Bruijne}, {Arenou}, {Babusiaux}, {Biermann}, {Creevey}, {Ducourant}, \& et~al.}]{GaiaDR3}
---, 2023, \aap, 674, A1

\bibitem[{{Garc{\'\i}a P{\'e}rez} {et~al}\mbox{.}(2016){Garc{\'\i}a P{\'e}rez}, {Allende Prieto}, {Holtzman}, {Shetrone}, {M{\'e}sz{\'a}ros}, {Bizyaev}, {Carrera}, {Cunha}, {Garc{\'\i}a-Hern{\'a}ndez}, {Johnson}, {Majewski}, {Nidever}, {Schiavon}, {Shane}, {Smith}, {Sobeck}, {Troup}, {Zamora}, {Weinberg}, {Bovy}, {Eisenstein}, {Feuillet}, {Frinchaboy}, {Hayden}, {Hearty}, {Nguyen}, {O'Connell}, {Pinsonneault}, {Wilson}, \& {Zasowski}}]{GarciaPerez2016}
{Garc{\'\i}a P{\'e}rez} A.~E. {et~al.}, 2016, \aj, 151, 144

\bibitem[{{Gonz{\'a}lez Hern{\'a}ndez} \& {Bonifacio}(2009)}]{GonzalezHernandez2009}
{Gonz{\'a}lez Hern{\'a}ndez} J.~I., {Bonifacio} P., 2009, \aap, 497, 497

\bibitem[{{Gunn} {et~al}\mbox{.}(2006){Gunn}, {Siegmund}, {Mannery}, {Owen}, {Hull}, {Leger}, {Carey}, {Knapp}, {York}, {Boroski}, {Kent}, {Lupton}, {Rockosi}, {Evans}, {Waddell}, {Anderson}, {Annis}, {Barentine}, {Bartoszek}, {Bastian}, {Bracker}, {Brewington}, {Briegel}, {Brinkmann}, {Brown}, {Carr}, {Czarapata}, {Drennan}, {Dombeck}, {Federwitz}, {Gillespie}, {Gonzales}, {Hansen}, {Harvanek}, {Hayes}, {Jordan}, {Kinney}, {Klaene}, {Kleinman}, {Kron}, {Kresinski}, {Lee}, {Limmongkol}, {Lindenmeyer}, {Long}, {Loomis}, {McGehee}, {Mantsch}, {Neilsen}, {Neswold}, {Newman}, {Nitta}, {Peoples}, {Pier}, {Prieto}, {Prosapio}, {Rivetta}, {Schneider}, {Snedden}, \& {Wang}}]{Gunn2006AJ}
{Gunn} J.~E. {et~al.}, 2006, \aj, 131, 2332

\bibitem[{{Gustafsson} {et~al}\mbox{.}(2008){Gustafsson}, {Edvardsson}, {Eriksson}, {J{\o}rgensen}, {Nordlund}, \& {Plez}}]{Gustafsson08}
{Gustafsson} B., {Edvardsson} B., {Eriksson} K., {J{\o}rgensen} U.~G., {Nordlund} {\r{A}}., {Plez} B., 2008, \aap, 486, 951

\bibitem[{{Hafen} {et~al}\mbox{.}(2022){Hafen}, {Stern}, {Bullock}, {Gurvich}, {Yu}, {Faucher-Gigu{\`e}re}, {Fielding}, {Angl{\'e}s-Alc{\'a}zar}, {Quataert}, {Wetzel}, {Starkenburg}, {Boylan-Kolchin}, {Moreno}, {Feldmann}, {El-Badry}, {Chan}, {Trapp}, {Kere{\v{s}}}, \& {Hopkins}}]{Hafen2022}
{Hafen} Z. {et~al.}, 2022, \mnras, 514, 5056

\bibitem[{{Hayes} {et~al}\mbox{.}(2023){Hayes}, {Venn}, {Waller}, {Jensen}, {McConnachie}, {Pazder}, {Sestito}, {Anthony}, {Baker}, {Bassett}, {Bento}, {Burley}, {Brzeski}, {Case}, {Chapin}, {Chin}, {Chisholm}, {Churilov}, {Densmore}, {Diaz}, {Dunn}, {Edgar}, {Farrell}, {Firpo}, {Fitzsimmons}, {Font-Serra}, {Fuentes}, {Ganton}, {Gomez-Jimenez}, {Hardy}, {Henderson}, {Hill}, {Hoff}, {Ireland}, {Kalari}, {Kelly}, {Klauser}, {Kondrat}, {Labrie}, {Lambert}, {Luvaul}, {Lawrence}, {Lothrop}, {Macdonald}, {Mali}, {Margheim}, {McDermid}, {McGregor}, {Miller}, {Miranda}, {Muller}, {Nielsen}, {Norbury}, {Oberdorf}, {Pai}, {Perez}, {Prado}, {Price}, {Quiroz}, {Reshetov}, {Robertson}, {Ruiz-Carmona}, {Salinas}, {Sebo}, {Sheinis}, {Shetrone}, {Shortridge}, {Silversides}, {Silva}, {Simpson}, {Smith}, {Szeto}, {Tims}, {Toro}, {Urrutia}, {Venkatesan}, {Waller}, {Wevers}, {Wierzbicki}, {White}, {Young}, \& {Zhelem}}]{Hayes23}
{Hayes} C.~R. {et~al.}, 2023, arXiv e-prints, arXiv:2306.04804

\bibitem[{{Hayes} {et~al}\mbox{.}(2022){Hayes}, {Waller}, {Ireland}, {Nielsen}, {White}, {Bento}, {Venn}, {Pazder}, {McConnachie}, {Simpson}, \& {Labrie}}]{Hayes22}
---, 2022, in Society of Photo-Optical Instrumentation Engineers (SPIE) Conference Series, Vol. 12184, Ground-based and Airborne Instrumentation for Astronomy IX, {Evans} C.~J., {Bryant} J.~J., {Motohara} K., eds., p. 121846H

\bibitem[{{Heger} {et~al}\mbox{.}(2012){Heger}, {Woosley}, {Vo}, {Chen}, \& {Joggerst}}]{Heger12}
{Heger} A., {Woosley} S., {Vo} P., {Chen} K., {Joggerst} C., 2012, in Astronomical Society of the Pacific Conference Series, Vol. 458, Galactic Archaeology: Near-Field Cosmology and the Formation of the Milky Way, {Aoki} W., {Ishigaki} M., {Suda} T., {Tsujimoto} T., {Arimoto} N., eds., p.~11

\bibitem[{{Heger} \& {Woosley}(2010)}]{Heger10}
{Heger} A., {Woosley} S.~E., 2010, \apj, 724, 341

\bibitem[{{Heiter} {et~al}\mbox{.}(2015){Heiter}, {Jofr{\'e}}, {Gustafsson}, {Korn}, {Soubiran}, \& {Th{\'e}venin}}]{Heiter2015}
{Heiter} U., {Jofr{\'e}} P., {Gustafsson} B., {Korn} A.~J., {Soubiran} C., {Th{\'e}venin} F., 2015, \aap, 582, A49

\bibitem[{{Helmi} {et~al}\mbox{.}(2018){Helmi}, {Babusiaux}, {Koppelman}, {Massari}, {Veljanoski}, \& {Brown}}]{Helmi18}
{Helmi} A., {Babusiaux} C., {Koppelman} H.~H., {Massari} D., {Veljanoski} J., {Brown} A. G.~A., 2018, \nat, 563, 85

\bibitem[{{Hill} {et~al}\mbox{.}(2019){Hill}, {Sk{\'u}lad{\'o}ttir}, {Tolstoy}, {Venn}, {Shetrone}, {Jablonka}, {Primas}, {Battaglia}, {de Boer}, {Fran{\c{c}}ois}, {Helmi}, {Kaufer}, {Letarte}, {Starkenburg}, \& {Spite}}]{Hill2019}
{Hill} V. {et~al.}, 2019, \aap, 626, A15

\bibitem[{{Ireland} {et~al}\mbox{.}(2012){Ireland}, {Barnes}, {Cochrane}, {Colless}, {Connor}, {Horton}, {Gibson}, {Lawrence}, {Martell}, {McGregor}, {Nicolle}, {Nield}, {Orr}, {Robertson}, {Ryder}, {Sheinis}, {Smith}, {Staszak}, {Tims}, {Xavier}, {Young}, \& {Zheng}}]{Ireland2012}
{Ireland} M.~J. {et~al.}, 2012, in Society of Photo-Optical Instrumentation Engineers (SPIE) Conference Series, Vol. 8446, Ground-based and Airborne Instrumentation for Astronomy IV, {McLean} I.~S., {Ramsay} S.~K., {Takami} H., eds., p. 844629

\bibitem[{{Ireland} {et~al}\mbox{.}(2018){Ireland}, {White}, {Bento}, {Farrell}, {Labrie}, {Luvaul}, {Nielsen}, \& {Simpson}}]{Ireland18}
{Ireland} M.~J., {White} M., {Bento} J.~P., {Farrell} T., {Labrie} K., {Luvaul} L., {Nielsen} J.~G., {Simpson} C., 2018, in Society of Photo-Optical Instrumentation Engineers (SPIE) Conference Series, Vol. 10707, Software and Cyberinfrastructure for Astronomy V, {Guzman} J.~C., {Ibsen} J., eds., p. 1070735

\bibitem[{{Ishigaki} {et~al}\mbox{.}(2014){Ishigaki}, {Aoki}, {Arimoto}, \& {Okamoto}}]{Ishigaki14}
{Ishigaki} M.~N., {Aoki} W., {Arimoto} N., {Okamoto} S., 2014, \aap, 562, A146

\bibitem[{{Ishigaki} {et~al}\mbox{.}(2018){Ishigaki}, {Tominaga}, {Kobayashi}, \& {Nomoto}}]{Ishigaki18}
{Ishigaki} M.~N., {Tominaga} N., {Kobayashi} C., {Nomoto} K., 2018, \apj, 857, 46

\bibitem[{{Jablonka} {et~al}\mbox{.}(2015){Jablonka}, {North}, {Mashonkina}, {Hill}, {Revaz}, {Shetrone}, {Starkenburg}, {Irwin}, {Tolstoy}, {Battaglia}, {Venn}, {Helmi}, {Primas}, \& {Fran{\c{c}}ois}}]{Jablonka15}
{Jablonka} P. {et~al.}, 2015, \aap, 583, A67

\bibitem[{{Ji} {et~al}\mbox{.}(2019){Ji}, {Simon}, {Frebel}, {Venn}, \& {Hansen}}]{Ji19}
{Ji} A.~P., {Simon} J.~D., {Frebel} A., {Venn} K.~A., {Hansen} T.~T., 2019, \apj, 870, 83

\bibitem[{{Johnston} {et~al}\mbox{.}(2008){Johnston}, {Bullock}, {Sharma}, {Font}, {Robertson}, \& {Leitner}}]{Johnston2008}
{Johnston} K.~V., {Bullock} J.~S., {Sharma} S., {Font} A., {Robertson} B.~E., {Leitner} S.~N., 2008, \apj, 689, 936

\bibitem[{{Karovicova} {et~al}\mbox{.}(2020){Karovicova}, {White}, {Nordlander}, {Casagrande}, {Ireland}, {Huber}, \& {Jofr{\'e}}}]{Karovicova20}
{Karovicova} I., {White} T.~R., {Nordlander} T., {Casagrande} L., {Ireland} M., {Huber} D., {Jofr{\'e}} P., 2020, \aap, 640, A25

\bibitem[{{Kielty} {et~al}\mbox{.}(2021){Kielty}, {Venn}, {Sestito}, {Starkenburg}, {Martin}, {Aguado}, {Arentsen}, {Fabbro}, {Gonz{\'a}lez Hern{\'a}ndez}, {Hill}, {Jablonka}, {Lardo}, {Mashonkina}, {Navarro}, {Sneden}, {Thomas}, {Youakim}, {Bialek}, \& {S{\'a}nchez-Janssen}}]{Kielty21}
{Kielty} C.~L. {et~al.}, 2021, \mnras, 506, 1438

\bibitem[{{Kraft} \& {Ivans}(2003)}]{Kraft2003}
{Kraft} R.~P., {Ivans} I.~I., 2003, \pasp, 115, 143

\bibitem[{{Labrie} {et~al}\mbox{.}(2022){Labrie}, {Simpson}, {Anderson}, {Cardenas}, {Turner}, {Quint}, {Conseil}, \& {Oberdorf}}]{DRAGONS2022}
{Labrie} K., {Simpson} C., {Anderson} K., {Cardenas} R., {Turner} J., {Quint} B., {Conseil} S., {Oberdorf} O., 2022, {DRAGONS}. Zenodo

\bibitem[{{Li} {et~al}\mbox{.}(2021){Li}, {Hammer}, {Babusiaux}, {Pawlowski}, {Yang}, {Arenou}, {Du}, \& {Wang}}]{LiH21}
{Li} H., {Hammer} F., {Babusiaux} C., {Pawlowski} M.~S., {Yang} Y., {Arenou} F., {Du} C., {Wang} J., 2021, \apj, 916, 8

\bibitem[{{Li} {et~al}\mbox{.}(2022){Li}, {Ji}, {Pace}, {Erkal}, {Koposov}, {Shipp}, {Da Costa}, {Cullinane}, {Kuehn}, {Lewis}, {Mackey}, {Simpson}, {Zucker}, {Ferguson}, {Martell}, {Bland-Hawthorn}, {Balbinot}, {Tavangar}, {Drlica-Wagner}, {De Silva}, \& {Simon}}]{Li22}
{Li} T.~S. {et~al.}, 2022, \apj, 928, 30

\bibitem[{{Lind} {et~al}\mbox{.}(2011){Lind}, {Asplund}, {Barklem}, \& {Belyaev}}]{Lind2011}
{Lind} K., {Asplund} M., {Barklem} P.~S., {Belyaev} A.~K., 2011, A\&A, 528, A103

\bibitem[{{Lind}, {Bergemann} \& {Asplund}(2012){Lind}, {Bergemann}, \& {Asplund}}]{Lind2012}
{Lind} K., {Bergemann} M., {Asplund} M., 2012, \mnras, 427, 50

\bibitem[{{Lindegren} {et~al}\mbox{.}(2021){Lindegren}, {Klioner}, {Hern{\'a}ndez}, {Bombrun}, {Ramos-Lerate}, {Steidelm{\"u}ller}, {Bastian}, {Biermann}, {de Torres}, {Gerlach}, {Geyer}, {Hilger}, {Hobbs}, {Lammers}, {McMillan}, {Stephenson}, {Casta{\~n}eda}, {Davidson}, {Fabricius}, {Gracia-Abril}, {Portell}, {Rowell}, {Teyssier}, {Torra}, {Bartolom{\'e}}, {Clotet}, {Garralda}, {Gonz{\'a}lez-Vidal}, {Torra}, {Abbas}, {Altmann}, {Anglada Varela}, {Balaguer-N{\'u}{\~n}ez}, {Balog}, {Barache}, {Becciani}, {Bernet}, {Bertone}, {Bianchi}, {Bouquillon}, {Brown}, {Bucciarelli}, {Busonero}, {Butkevich}, {Buzzi}, {Cancelliere}, {Carlucci}, {Charlot}, {Cioni}, {Crosta}, {Crowley}, {del Peloso}, {del Pozo}, {Drimmel}, {Esquej}, {Fienga}, {Fraile}, {Gai}, {Garcia-Reinaldos}, {Guerra}, {Hambly}, {Hauser}, {Jan{\ss}en}, {Jordan}, {Kostrzewa-Rutkowska}, {Lattanzi}, {Liao}, {Licata}, {Lister}, {L{\"o}ffler}, {Marchant}, {Masip}, {Mignard}, {Mints}, {Molina}, {Mora}, {Morbidelli}, {Murphy}, {Pagani}, {Panuzzo},
  {Pe{\~n}alosa Esteller}, {Poggio}, {Re Fiorentin}, {Riva}, {Sagrist{\`a} Sell{\'e}s}, {Sanchez Gimenez}, {Sarasso}, {Sciacca}, {Siddiqui}, {Smart}, {Souami}, {Spagna}, {Steele}, {Taris}, {Utrilla}, {van Reeven}, \& {Vecchiato}}]{Lindegren21}
{Lindegren} L. {et~al.}, 2021, \aap, 649, A2

\bibitem[{{Lucchesi} {et~al}\mbox{.}(2022){Lucchesi}, {Lardo}, {Jablonka}, {Sestito}, {Mashonkina}, {Arentsen}, {Suter}, {Venn}, {Martin}, {Starkenburg}, {Aguado}, {Hill}, {Kordopatis}, {Navarro}, {Gonz{\'a}lez Hern{\'a}ndez}, {Malhan}, \& {Yuan}}]{Lucchesi22}
{Lucchesi} R. {et~al.}, 2022, \mnras, 511, 1004

\bibitem[{{Majewski} {et~al}\mbox{.}(2017){Majewski}, {Schiavon}, {Frinchaboy}, {Allende Prieto}, {Barkhouser}, {Bizyaev}, {Blank}, {Brunner}, {Burton}, {Carrera}, {Chojnowski}, {Cunha}, {Epstein}, {Fitzgerald}, {Garc{\'\i}a P{\'e}rez}, {Hearty}, {Henderson}, {Holtzman}, {Johnson}, {Lam}, {Lawler}, {Maseman}, {M{\'e}sz{\'a}ros}, {Nelson}, {Nguyen}, {Nidever}, {Pinsonneault}, {Shetrone}, {Smee}, {Smith}, {Stolberg}, {Skrutskie}, {Walker}, {Wilson}, {Zasowski}, {Anders}, {Basu}, {Beland}, {Blanton}, {Bovy}, {Brownstein}, {Carlberg}, {Chaplin}, {Chiappini}, {Eisenstein}, {Elsworth}, {Feuillet}, {Fleming}, {Galbraith-Frew}, {Garc{\'\i}a}, {Garc{\'\i}a-Hern{\'a}ndez}, {Gillespie}, {Girardi}, {Gunn}, {Hasselquist}, {Hayden}, {Hekker}, {Ivans}, {Kinemuchi}, {Klaene}, {Mahadevan}, {Mathur}, {Mosser}, {Muna}, {Munn}, {Nichol}, {O'Connell}, {Parejko}, {Robin}, {Rocha-Pinto}, {Schultheis}, {Serenelli}, {Shane}, {Silva Aguirre}, {Sobeck}, {Thompson}, {Troup}, {Weinberg}, \& {Zamora}}]{apogee2017}
{Majewski} S.~R. {et~al.}, 2017, \aj, 154, 94

\bibitem[{{Mardini} {et~al}\mbox{.}(2022{\natexlab{a}}){Mardini}, {Frebel}, {Chiti}, {Meiron}, {Brauer}, \& {Ou}}]{Mardini22b}
{Mardini} M.~K., {Frebel} A., {Chiti} A., {Meiron} Y., {Brauer} K.~V., {Ou} X., 2022{\natexlab{a}}, \apj, 936, 78

\bibitem[{{Mardini} {et~al}\mbox{.}(2022{\natexlab{b}}){Mardini}, {Frebel}, {Ezzeddine}, {Chiti}, {Meiron}, {Ji}, {Placco}, {Roederer}, \& {Mel{\'e}ndez}}]{Mardini22}
{Mardini} M.~K. {et~al.}, 2022{\natexlab{b}}, \mnras, 517, 3993

\bibitem[{{Martin} {et~al}\mbox{.}(2023){Martin}, {Starkenburg}, {Yuan}, {Fouesneau}, {Arentsen}, {De Angeli}, {Gran}, {Montelius}, {Andrae}, {Bellazzini}, {Montegriffo}, {Esselink}, {Zhang}, {Venn}, {Viswanathan}, {Aguado}, {Battaglia}, {Bayer}, {Bonifacio}, {Caffau}, {C{\^o}t{\'e}}, {Carlberg}, {Fabbro}, {Fern{\'a}ndez Alvar}, {Gonz{\'a}lez Hern{\'a}ndez}, {Gonz{\'a}lez Rivera de La Vernhe}, {Hill}, {Ibata}, {Jablonka}, {Kordopatis}, {Lardo}, {McConnachie}, {Navarrete}, {Navarro}, {Recio-Blanco}, {S{\'a}nchez Janssen}, {Sestito}, {Thomas}, {Vitali}, \& {Youakim}}]{Martin2023}
{Martin} N.~F. {et~al.}, 2023, arXiv e-prints, arXiv:2308.01344

\bibitem[{{Mashonkina} {et~al}\mbox{.}(2017){Mashonkina}, {Jablonka}, {Pakhomov}, {Sitnova}, \& {North}}]{Mashonkina17}
{Mashonkina} L., {Jablonka} P., {Pakhomov} Y., {Sitnova} T., {North} P., 2017, \aap, 604, A129

\bibitem[{{Masseron} {et~al}\mbox{.}(2014){Masseron}, {Plez}, {Van Eck}, {Colin}, {Daoutidis}, {Godefroid}, {Coheur}, {Bernath}, {Jorissen}, \& {Christlieb}}]{Masseron2014}
{Masseron} T. {et~al.}, 2014, \aap, 571, A47

\bibitem[{{McConnachie} {et~al}\mbox{.}(2022){McConnachie}, {Hayes}, {Ireland}, {Waller}, {Berg}, {Pazder}, {Margheim}, {Kalari}, {Farrell}, \& {Robertson}}]{McConnachie2022}
{McConnachie} A.~W. {et~al.}, 2022, in Society of Photo-Optical Instrumentation Engineers (SPIE) Conference Series, Vol. 12184, Ground-based and Airborne Instrumentation for Astronomy IX, {Evans} C.~J., {Bryant} J.~J., {Motohara} K., eds., p. 121841E

\bibitem[{{Minchev} \& {Famaey}(2010)}]{Minchev10}
{Minchev} I., {Famaey} B., 2010, \apj, 722, 112

\bibitem[{{Mucciarelli}, {Bellazzini} \& {Massari}(2021){Mucciarelli}, {Bellazzini}, \& {Massari}}]{Mucciarelli21}
{Mucciarelli} A., {Bellazzini} M., {Massari} D., 2021, \aap, 653, A90

\bibitem[{{Navarro} {et~al}\mbox{.}(2018){Navarro}, {Yozin}, {Loewen}, {Ben{\'\i}tez-Llambay}, {Fattahi}, {Frenk}, {Oman}, {Schaye}, \& {Theuns}}]{Navarro2018}
{Navarro} J.~F. {et~al.}, 2018, \mnras, 476, 3648

\bibitem[{{Nidever} {et~al}\mbox{.}(2015){Nidever}, {Holtzman}, {Allende Prieto}, {Beland}, {Bender}, {Bizyaev}, {Burton}, {Desphande}, {Fleming}, {Garc{\'\i}a P{\'e}rez}, {Hearty}, {Majewski}, {M{\'e}sz{\'a}ros}, {Muna}, {Nguyen}, {Schiavon}, {Shetrone}, {Skrutskie}, {Sobeck}, \& {Wilson}}]{Nidever2015}
{Nidever} D.~L. {et~al.}, 2015, \aj, 150, 173

\bibitem[{{Nomoto}, {Kobayashi} \& {Tominaga}(2013){Nomoto}, {Kobayashi}, \& {Tominaga}}]{Nomoto13}
{Nomoto} K., {Kobayashi} C., {Tominaga} N., 2013, \araa, 51, 457

\bibitem[{{Nordlander} {et~al}\mbox{.}(2017){Nordlander}, {Amarsi}, {Lind}, {Asplund}, {Barklem}, {Casey}, {Collet}, \& {Leenaarts}}]{Nordlander17}
{Nordlander} T., {Amarsi} A.~M., {Lind} K., {Asplund} M., {Barklem} P.~S., {Casey} A.~R., {Collet} R., {Leenaarts} J., 2017, \aap, 597, A6

\bibitem[{{Nordlander} \& {Lind}(2017)}]{Nordlander17a}
{Nordlander} T., {Lind} K., 2017, \aap, 607, A75

\bibitem[{{Pazder} {et~al}\mbox{.}(2020){Pazder}, {McConnachie}, {Ireland}, {Anthony}, {Bassett}, {Burley}, {Chapin}, {Churilov}, {Densmore}, {Dunn}, {Farrell}, {Henderson}, {Hoff}, {Lambert}, {Lothrop}, {MacDonald}, {Margheim}, {Reshetov}, {Wevers}, {Waller}, {Young}, \& {Zhelem}}]{Pazder20}
{Pazder} J. {et~al.}, 2020, in Society of Photo-Optical Instrumentation Engineers (SPIE) Conference Series, Vol. 11447, Society of Photo-Optical Instrumentation Engineers (SPIE) Conference Series, p. 1144743

\bibitem[{{Placco} {et~al}\mbox{.}(2016){Placco}, {Beers}, {Reggiani}, \& {Mel{\'e}ndez}}]{Placco2016}
{Placco} V.~M., {Beers} T.~C., {Reggiani} H., {Mel{\'e}ndez} J., 2016, \apjl, 829, L24

\bibitem[{{Placco} {et~al}\mbox{.}(2020){Placco}, {Santucci}, {Yuan}, {Mardini}, {Holmbeck}, {Wang}, {Surman}, {Hansen}, {Roederer}, {Beers}, {Choplin}, {Ji}, {Ezzeddine}, {Frebel}, {Sakari}, {Whitten}, \& {Zepeda}}]{Placco2020}
{Placco} V.~M. {et~al.}, 2020, \apj, 897, 78

\bibitem[{{Placco} {et~al}\mbox{.}(2021){Placco}, {Sneden}, {Roederer}, {Lawler}, {Den Hartog}, {Hejazi}, {Maas}, \& {Bernath}}]{Placco21}
{Placco} V.~M., {Sneden} C., {Roederer} I.~U., {Lawler} J.~E., {Den Hartog} E.~A., {Hejazi} N., {Maas} Z., {Bernath} P., 2021, Research Notes of the American Astronomical Society, 5, 92

\bibitem[{{Queiroz} {et~al}\mbox{.}(2020){Queiroz}, {Anders}, {Chiappini}, {Khalatyan}, {Santiago}, {Steinmetz}, {Valentini}, {Miglio}, {Bossini}, {Barbuy}, {Minchev}, {Minniti}, {Garc{\'\i}a Hern{\'a}ndez}, {Schultheis}, {Beaton}, {Beers}, {Bizyaev}, {Brownstein}, {Cunha}, {Fern{\'a}ndez-Trincado}, {Frinchaboy}, {Lane}, {Majewski}, {Nataf}, {Nitschelm}, {Pan}, {Roman-Lopes}, {Sobeck}, {Stringfellow}, \& {Zamora}}]{Queiroz2020}
{Queiroz} A.~B.~A. {et~al.}, 2020, \aap, 638, A76

\bibitem[{{Sales} {et~al}\mbox{.}(2012){Sales}, {Navarro}, {Theuns}, {Schaye}, {White}, {Frenk}, {Crain}, \& {Dalla Vecchia}}]{Sales2012}
{Sales} L.~V., {Navarro} J.~F., {Theuns} T., {Schaye} J., {White} S. D.~M., {Frenk} C.~S., {Crain} R.~A., {Dalla Vecchia} C., 2012, \mnras, 423, 1544

\bibitem[{{Santana} {et~al}\mbox{.}(2021){Santana}, {Beaton}, {Covey}, {O'Connell}, {Longa-Pe{\~n}a}, {Cohen}, {Fern{\'a}ndez-Trincado}, {Hayes}, {Zasowski}, {Sobeck}, {Majewski}, {Chojnowski}, {De Lee}, {Oelkers}, {Stringfellow}, {Almeida}, {Anguiano}, {Donor}, {Frinchaboy}, {Hasselquist}, {Johnson}, {Kollmeier}, {Nidever}, {Price-Whelan}, {Rojas-Arriagada}, {Schultheis}, {Shetrone}, {Simon}, {Aerts}, {Borissova}, {Drout}, {Geisler}, {Law}, {Medina}, {Minniti}, {Monachesi}, {Mu{\~n}oz}, {Poleski}, {Roman-Lopes}, {Schlaufman}, {Stutz}, {Teske}, {Tkachenko}, {Van Saders}, {Weinberger}, \& {Zoccali}}]{Santana2021}
{Santana} F.~A. {et~al.}, 2021, \aj, 162, 303

\bibitem[{{Santistevan} {et~al}\mbox{.}(2021){Santistevan}, {Wetzel}, {Sanderson}, {El-Badry}, {Samuel}, \& {Faucher-Gigu{\`e}re}}]{Santistevan21}
{Santistevan} I.~B., {Wetzel} A., {Sanderson} R.~E., {El-Badry} K., {Samuel} J., {Faucher-Gigu{\`e}re} C.-A., 2021, \mnras, 505, 921

\bibitem[{{Scannapieco} {et~al}\mbox{.}(2011){Scannapieco}, {White}, {Springel}, \& {Tissera}}]{Scannapieco11}
{Scannapieco} C., {White} S.~D.~M., {Springel} V., {Tissera} P.~B., 2011, \mnras, 417, 154

\bibitem[{{Schlafly} \& {Finkbeiner}(2011)}]{Schlafly11}
{Schlafly} E.~F., {Finkbeiner} D.~P., 2011, \apj, 737, 103

\bibitem[{{Schlaufman} {et~al}\mbox{.}(2018){Schlaufman}, {Thompson}, {Casey}, \& {}}]{Schlaufman2018}
{Schlaufman} K.~C., {Thompson} I.~B., {Casey} A.~R., {}, 2018, \apj, 867, 98

\bibitem[{{Sestito} {et~al}\mbox{.}(2021){Sestito}, {Buck}, {Starkenburg}, {Martin}, {Navarro}, {Venn}, {Obreja}, {Jablonka}, \& {Macci{\`o}}}]{Sestito21}
{Sestito} F. {et~al.}, 2021, \mnras, 500, 3750

\bibitem[{{Sestito} {et~al}\mbox{.}(2023{\natexlab{a}}){Sestito}, {Hayes}, {Venn}, {Jensen}, {McConnachie}, {Pazder}, {Waller}, {Arentsen}, {Jablonka}, {Martin}, {Matsuno}, {Navarro}, {Starkenburg}, {Vitali}, {Bassett}, {Diaz}, {Edgar}, {Firpo}, {Gomez-Jimenez}, {Kalari}, {Lambert}, {Lawrence}, {Robertson}, {Ruiz-Carmona}, {Salinas}, {Sebo}, \& {Venkatesan}}]{Sestito23gh}
---, 2023{\natexlab{a}}, arXiv e-prints, arXiv:2308.07366

\bibitem[{{Sestito} {et~al}\mbox{.}(2019){Sestito}, {Longeard}, {Martin}, {Starkenburg}, {Fouesneau}, {Gonz{\'a}lez Hern{\'a}ndez}, {Arentsen}, {Ibata}, {Aguado}, {Carlberg}, {Jablonka}, {Navarro}, {Tolstoy}, \& {Venn}}]{Sestito19}
---, 2019, \mnras, 484, 2166

\bibitem[{{Sestito} {et~al}\mbox{.}(2020){Sestito}, {Martin}, {Starkenburg}, {Arentsen}, {Ibata}, {Longeard}, {Kielty}, {Youakim}, {Venn}, {Aguado}, {Carlberg}, {Gonz{\'a}lez Hern{\'a}ndez}, {Hill}, {Jablonka}, {Kordopatis}, {Malhan}, {Navarro}, {S{\'a}nchez-Janssen}, {Thomas}, {Tolstoy}, {Wilson}, {Palicio}, {Bialek}, {Garcia-Dias}, {Lucchesi}, {North}, {Osorio}, {Patrick}, \& {Peralta de Arriba}}]{Sestito20}
---, 2020, \mnras, 497, L7

\bibitem[{{Sestito} {et~al}\mbox{.}(2023{\natexlab{b}}){Sestito}, {Venn}, {Arentsen}, {Aguado}, {Kielty}, {Lardo}, {Martin}, {Navarro}, {Starkenburg}, {Waller}, {Carlberg}, {Fran{\c{c}}ois}, {Gonz{\'a}lez Hern{\'a}ndez}, {Kordopatis}, {Vitali}, \& {Yuan}}]{Sestito23}
---, 2023{\natexlab{b}}, \mnras, 518, 4557

\bibitem[{{Shetrone} {et~al}\mbox{.}(2015){Shetrone}, {Bizyaev}, {Lawler}, {Allende Prieto}, {Johnson}, {Smith}, {Cunha}, {Holtzman}, {Garc{\'\i}a P{\'e}rez}, {M{\'e}sz{\'a}ros}, {Sobeck}, {Zamora}, {Garc{\'\i}a-Hern{\'a}ndez}, {Souto}, {Chojnowski}, {Koesterke}, {Majewski}, \& {Zasowski}}]{Shetrone2015}
{Shetrone} M. {et~al.}, 2015, \apjs, 221, 24

\bibitem[{{Sitnova} {et~al}\mbox{.}(2015){Sitnova}, {Zhao}, {Mashonkina}, {Chen}, {Liu}, {Pakhomov}, {Tan}, {Bolte}, {Alexeeva}, {Grupp}, {Shi}, \& {Zhang}}]{Sitnova2015}
{Sitnova} T. {et~al.}, 2015, \apj, 808, 148

\bibitem[{{Sitnova} {et~al}\mbox{.}(2021){Sitnova}, {Mashonkina}, {Tatarnikov}, {Voziakova}, {Burlak}, {Pakhomov}, {Jablonka}, {Neretina}, \& {Frebel}}]{Sitnova21}
{Sitnova} T.~M. {et~al.}, 2021, \mnras, 504, 1183

\bibitem[{{Sk{\'u}lad{\'o}ttir} {et~al}\mbox{.}(2021){Sk{\'u}lad{\'o}ttir}, {Salvadori}, {Amarsi}, {Tolstoy}, {Irwin}, {Hill}, {Jablonka}, {Battaglia}, {Starkenburg}, {Massari}, {Helmi}, \& {Posti}}]{Skuladottir21}
{Sk{\'u}lad{\'o}ttir} {\'A}. {et~al.}, 2021, \apjl, 915, L30

\bibitem[{{Sk{\'u}lad{\'o}ttir} {et~al}\mbox{.}(2023){Sk{\'u}lad{\'o}ttir}, {Vanni}, {Salvadori}, \& {Lucchesi}}]{Skuladottir2023}
{Sk{\'u}lad{\'o}ttir} {\'A}., {Vanni} I., {Salvadori} S., {Lucchesi} R., 2023, arXiv e-prints, arXiv:2305.02829

\bibitem[{{Smith} {et~al}\mbox{.}(2021){Smith}, {Bizyaev}, {Cunha}, {Shetrone}, {Souto}, {Allende Prieto}, {Masseron}, {M{\'e}sz{\'a}ros}, {J{\"o}nsson}, {Hasselquist}, {Osorio}, {Garc{\'\i}a-Hern{\'a}ndez}, {Plez}, {Beaton}, {Holtzman}, {Majewski}, {Stringfellow}, \& {Sobeck}}]{Smith2021}
{Smith} V.~V. {et~al.}, 2021, \aj, 161, 254

\bibitem[{{Sneden}(1973)}]{Sneden73}
{Sneden} C.~A., 1973, PhD thesis, THE UNIVERSITY OF TEXAS AT AUSTIN.

\bibitem[{{Sobeck} {et~al}\mbox{.}(2011){Sobeck}, {Kraft}, {Sneden}, {Preston}, {Cowan}, {Smith}, {Thompson}, {Shectman}, \& {Burley}}]{Sobeck11}
{Sobeck} J.~S. {et~al.}, 2011, \aj, 141, 175

\bibitem[{{Spite}, {Spite} \& {Barbuy}(2021){Spite}, {Spite}, \& {Barbuy}}]{Spite2021}
{Spite} M., {Spite} F., {Barbuy} B., 2021, \aap, 652, A97

\bibitem[{{Starkenburg} {et~al}\mbox{.}(2017{\natexlab{a}}){Starkenburg}, {Martin}, {Youakim}, {Aguado}, {Allende Prieto}, {Arentsen}, {Bernard}, {Bonifacio}, {Caffau}, {Carlberg}, {C{\^o}t{\'e}}, {Fouesneau}, {Fran{\c c}ois}, {Franke}, {Gonz{\'a}lez Hern{\'a}ndez}, {Gwyn}, {Hill}, {Ibata}, {Jablonka}, {Longeard}, {McConnachie}, {Navarro}, {S{\'a}nchez-Janssen}, {Tolstoy}, \& {Venn}}]{Starkenburg17b}
{Starkenburg} E. {et~al.}, 2017{\natexlab{a}}, \mnras, 471, 2587

\bibitem[{{Starkenburg} {et~al}\mbox{.}(2017{\natexlab{b}}){Starkenburg}, {Oman}, {Navarro}, {Crain}, {Fattahi}, {Frenk}, {Sawala}, \& {Schaye}}]{Starkenburg17a}
{Starkenburg} E., {Oman} K.~A., {Navarro} J.~F., {Crain} R.~A., {Fattahi} A., {Frenk} C.~S., {Sawala} T., {Schaye} J., 2017{\natexlab{b}}, \mnras, 465, 2212

\bibitem[{{Stern} {et~al}\mbox{.}(2021){Stern}, {Faucher-Gigu{\`e}re}, {Fielding}, {Quataert}, {Hafen}, {Gurvich}, {Ma}, {Byrne}, {El-Badry}, {Angl{\'e}s-Alc{\'a}zar}, {Chan}, {Feldmann}, {Kere{\v{s}}}, {Wetzel}, {Murray}, \& {Hopkins}}]{Stern2021}
{Stern} J. {et~al.}, 2021, \apj, 911, 88

\bibitem[{{Suda} {et~al}\mbox{.}(2017){Suda}, {Hidaka}, {Aoki}, {Katsuta}, {Yamada}, {Fujimoto}, {Ohtani}, {Masuyama}, {Noda}, \& {Wada}}]{Suda2017}
{Suda} T. {et~al.}, 2017, \pasj, 69, 76

\bibitem[{{Suda} {et~al}\mbox{.}(2008){Suda}, {Katsuta}, {Yamada}, {Suwa}, {Ishizuka}, {Komiya}, {Sorai}, {Aikawa}, \& {Fujimoto}}]{Suda08}
---, 2008, \pasj, 60, 1159

\bibitem[{{Taylor}(2005)}]{Taylor05}
{Taylor} M.~B., 2005, in Astronomical Society of the Pacific Conference Series, Vol. 347, Astronomical Data Analysis Software and Systems XIV, {Shopbell} P., {Britton} M., {Ebert} R., eds., p.~29

\bibitem[{{Tody}(1986)}]{Tody86}
{Tody} D., 1986, in Society of Photo-Optical Instrumentation Engineers (SPIE) Conference Series, Vol. 627, Society of Photo-Optical Instrumentation Engineers (SPIE), {Crawford} D.~L., ed., p. 733

\bibitem[{{Tody}(1993)}]{Tody93}
---, 1993, in Astronomical Society of the Pacific Conference Series, Vol.~52, Astronomical Data Analysis Software and Systems II, {Hanisch} R.~J., {Brissenden} R.~J.~V., {Barnes} J., eds., p. 173

\bibitem[{{Tolstoy}, {Hill} \& {Tosi}(2009){Tolstoy}, {Hill}, \& {Tosi}}]{Tolstoy09}
{Tolstoy} E., {Hill} V., {Tosi} M., 2009, \araa, 47, 371

\bibitem[{{Venn} {et~al}\mbox{.}(2004){Venn}, {Irwin}, {Shetrone}, {Tout}, {Hill}, \& {Tolstoy}}]{Venn04}
{Venn} K.~A., {Irwin} M., {Shetrone} M.~D., {Tout} C.~A., {Hill} V., {Tolstoy} E., 2004, \aj, 128, 1177

\bibitem[{{Venn} {et~al}\mbox{.}(2020){Venn}, {Kielty}, {Sestito}, {Starkenburg}, {Martin}, {Aguado}, {Arentsen}, {Bonifacio}, {Caffau}, {Hill}, {Jablonka}, {Lardo}, {Mashonkina}, {Navarro}, {Sneden}, {Thomas}, {Youakim}, {Gonz{\'a}lez-Hern{\'a}ndez}, {S{\'a}nchez Janssen}, {Carlberg}, \& {Malhan}}]{Venn20}
{Venn} K.~A. {et~al.}, 2020, \mnras, 492, 3241

\bibitem[{{Venn} {et~al}\mbox{.}(2012){Venn}, {Shetrone}, {Irwin}, {Hill}, {Jablonka}, {Tolstoy}, {Lemasle}, {Divell}, {Starkenburg}, {Letarte}, {Baldner}, {Battaglia}, {Helmi}, {Kaufer}, \& {Primas}}]{Venn12}
---, 2012, \apj, 751, 102

\bibitem[{{Venn} {et~al}\mbox{.}(2017){Venn}, {Starkenburg}, {Malo}, {Martin}, \& {Laevens}}]{Venn2017}
{Venn} K.~A., {Starkenburg} E., {Malo} L., {Martin} N., {Laevens} B.~P.~M., 2017, \mnras, 466, 3741

\bibitem[{{Waller} {et~al}\mbox{.}(2023){Waller}, {Venn}, {Sestito}, {Jensen}, {Kielty}, {Borukhovetskaya}, {Hayes}, {McConnachie}, \& {Navarro}}]{Waller23}
{Waller} F. {et~al.}, 2023, \mnras, 519, 1349

\bibitem[{{Wenger} {et~al}\mbox{.}(2000){Wenger}, {Ochsenbein}, {Egret}, {Dubois}, {Bonnarel}, {Borde}, {Genova}, {Jasniewicz}, {Lalo{\"e}}, {Lesteven}, \& {Monier}}]{Wenger00}
{Wenger} M. {et~al.}, 2000, \aaps, 143, 9

\bibitem[{{Wilson} {et~al}\mbox{.}(2019){Wilson}, {Hearty}, {Skrutskie}, {Majewski}, {Holtzman}, {Eisenstein}, {Gunn}, {Blank}, {Henderson}, {Smee}, {Nelson}, {Nidever}, {Arns}, {Barkhouser}, {Barr}, {Beland}, {Bershady}, {Blanton}, {Brunner}, {Burton}, {Carey}, {Carr}, {Colque}, {Crane}, {Damke}, {Davidson}, {Dean}, {Di Mille}, {Don}, {Ebelke}, {Evans}, {Fitzgerald}, {Gillespie}, {Hall}, {Harding}, {Harding}, {Hammond}, {Hancock}, {Harrison}, {Hope}, {Horne}, {Karakla}, {Lam}, {Leger}, {MacDonald}, {Maseman}, {Matsunari}, {Melton}, {Mitcheltree}, {O'Brien}, {O'Connell}, {Patten}, {Richardson}, {Rieke}, {Rieke}, {Roman-Lopes}, {Schiavon}, {Sobeck}, {Stolberg}, {Stoll}, {Tembe}, {Trujillo}, {Uomoto}, {Vernieri}, {Walker}, {Weinberg}, {Young}, {Anthony-Brumfield}, {Bizyaev}, {Breslauer}, {De Lee}, {Downey}, {Halverson}, {Huehnerhoff}, {Klaene}, {Leon}, {Long}, {Mahadevan}, {Malanushenko}, {Nguyen}, {Owen}, {S{\'a}nchez-Gallego}, {Sayres}, {Shane}, {Shectman}, {Shetrone}, {Skinner}, {Stauffer}, \&
  {Zhao}}]{Wilson2019}
{Wilson} J.~C. {et~al.}, 2019, \pasp, 131, 055001

\bibitem[{{Yong} {et~al}\mbox{.}(2021){Yong}, {Da Costa}, {Bessell}, {Chiti}, {Frebel}, {Gao}, {Lind}, {Mackey}, {Marino}, {Murphy}, {Nordlander}, {Asplund}, {Casey}, {Kobayashi}, {Norris}, \& {Schmidt}}]{Yong21}
{Yong} D. {et~al.}, 2021, \mnras, 507, 4102

\bibitem[{{Yong} {et~al}\mbox{.}(2013){Yong}, {Norris}, {Bessell}, {Christlieb}, {Asplund}, {Beers}, {Barklem}, {Frebel}, \& {Ryan}}]{Yong13}
---, 2013, \apj, 762, 26

\bibitem[{{Youakim} {et~al}\mbox{.}(2017){Youakim}, {Starkenburg}, {Aguado}, {Martin}, {Fouesneau}, {Gonz{\'a}lez Hern{\'a}ndez}, {Allende Prieto}, {Bonifacio}, {Gentile}, {Kielty}, {C{\^o}t{\'e}}, {Jablonka}, {McConnachie}, {S{\'a}nchez Janssen}, {Tolstoy}, \& {Venn}}]{Youakim17}
{Youakim} K. {et~al.}, 2017, \mnras, 472, 2963

\bibitem[{{Yu} {et~al}\mbox{.}(2021){Yu}, {Bullock}, {Klein}, {Stern}, {Wetzel}, {Ma}, {Moreno}, {Hafen}, {Gurvich}, {Hopkins}, {Kere{\v{s}}}, {Faucher-Gigu{\`e}re}, {Feldmann}, \& {Quataert}}]{Yu2021}
{Yu} S. {et~al.}, 2021, \mnras, 505, 889

\bibitem[{{Zasowski} {et~al}\mbox{.}(2017){Zasowski}, {Cohen}, {Chojnowski}, {Santana}, {Oelkers}, {Andrews}, {Beaton}, {Bender}, {Bird}, {Bovy}, {Carlberg}, {Covey}, {Cunha}, {Dell'Agli}, {Fleming}, {Frinchaboy}, {Garc{\'\i}a-Hern{\'a}ndez}, {Harding}, {Holtzman}, {Johnson}, {Kollmeier}, {Majewski}, {M{\'e}sz{\'a}ros}, {Munn}, {Mu{\~n}oz}, {Ness}, {Nidever}, {Poleski}, {Rom{\'a}n-Z{\'u}{\~n}iga}, {Shetrone}, {Simon}, {Smith}, {Sobeck}, {Stringfellow}, {Szigeti{\'a}ros}, {Tayar}, \& {Troup}}]{Zasowski2017}
{Zasowski} G. {et~al.}, 2017, \aj, 154, 198

\bibitem[{{Zasowski} {et~al}\mbox{.}(2013){Zasowski}, {Johnson}, {Frinchaboy}, {Majewski}, {Nidever}, {Rocha Pinto}, {Girardi}, {Andrews}, {Chojnowski}, {Cudworth}, {Jackson}, {Munn}, {Skrutskie}, {Beaton}, {Blake}, {Covey}, {Deshpande}, {Epstein}, {Fabbian}, {Fleming}, {Garcia Hernandez}, {Herrero}, {Mahadevan}, {M{\'e}sz{\'a}ros}, {Schultheis}, {Sellgren}, {Terrien}, {van Saders}, {Allende Prieto}, {Bizyaev}, {Burton}, {Cunha}, {da Costa}, {Hasselquist}, {Hearty}, {Holtzman}, {Garc{\'\i}a P{\'e}rez}, {Maia}, {O'Connell}, {O'Donnell}, {Pinsonneault}, {Santiago}, {Schiavon}, {Shetrone}, {Smith}, \& {Wilson}}]{Zasowski2013}
---, 2013, \aj, 146, 81

\end{thebibliography}

\section{Appendix A: Line Lists}

\clearpage

\begin{table}
\caption{Iron line list.  The source and quality of the atomic data is provided as Q, where A= OBR91 ($<$10\%), B= NIST C (25\%), C= NIST D+ (50\%) precision.  NLTE corrections  ($\Delta$) are from the MPIA database.}
\label{tab:lines_fe}
\begin{tabular}{ccccccc}
\hline
   $\lambda$ &  Elem &  $\chi$ &  log(gf) &   Q & A(X) & $\Delta$ \\
   (\AA) & & (eV) & & & LTE & NLTE \\
\hline

3758.233 &  26.0 & 0.957 &     -0.01 &  A & 4.07 & 0.13 \\
3763.789 &  26.0 & 0.989 &     -0.22 &  A & 4.22 & 0.14 \\
3787.880 &  26.0 & 1.010 &     -0.84 &  A & 4.37 & 0.14 \\
3815.840 &  26.0 & 1.484 &      0.24 &  A & 4.42 & 0.14\\
3820.425 &  26.0 & 0.858 &      0.16 &  A & 4.32 & 0.10\\
3824.444 &  26.0 & 0.000 &     -1.34 &  A & 4.72 & 0.14\\
3825.881 &  26.0 & 0.914 &     -0.02 &  A & 4.27 & 0.11\\
3827.823 &  26.0 & 1.556 &      0.09 &  A & 4.22 & 0.14\\
3840.438 &  26.0 & 0.989 &     -0.50 &  A & 4.22 & 0.14\\
3841.048 &  26.0 & 1.607 &     -0.04 &  A & 4.27 & 0.14\\
3849.967 &  26.0 & 1.010 &     -0.86 &  A & 4.27 & 0.12\\
3856.372 &  26.0 & 0.052 &     -1.28 &  A & 4.62 & 0.14\\
3859.912 &  26.0 & 0.000 &     -0.70 &  A & 4.77 & 0.12\\
3878.018 &  26.0 & 0.957 &     -0.90 &  A & 4.27 & 0.14\\
3878.573 &  26.0 & 0.087 &     -1.38 &  A & 4.47 & 0.14\\
3895.656 &  26.0 & 0.110 &     -1.67 &  A & 4.37 & 0.14\\
3902.946 &  26.0 & 1.556 &     -0.44 &  A & 4.22 & 0.14\\
3920.258 &  26.0 & 0.121 &     -1.73 &  A & 4.52 & 0.14\\
3922.912 &  26.0 & 0.052 &     -1.63 &  A & 4.42 & 0.15\\
4005.242 &  26.0 & 1.556 &     -0.58 &  A & 4.27 & 0.14\\
4045.812 &  26.0 & 1.484 &      0.28 &  A & 4.22 & 0.13\\
4063.594 &  26.0 & 1.556 &      0.06 &  A & 4.27 & 0.14\\
4071.738 &  26.0 & 1.607 &     -0.01 &  A & 4.27 & 0.14\\
4132.058 &  26.0 & 1.607 &     -0.68 &  A & 4.32 & 0.14\\
4143.868 &  26.0 & 1.556 &     -0.51 &  A & 4.17 & 0.14\\
4202.029 &  26.0 & 1.484 &     -0.69 &  A & 4.22 & 0.14\\
4250.787 &  26.0 & 1.556 &     -0.71 &  A & 4.27 & 0.14\\
4260.474 &  26.0 & 2.397 &     -0.02 &   $-$ & 4.27 & 0.13\\
4271.761 &  26.0 & 1.484 &     -0.17 &  A & 4.22 & 0.14\\
4282.403 &  26.0 & 2.174 &     -0.78 &  A & 4.52 & 0.14\\
4325.762 &  26.0 & 1.607 &      0.01 &  A & 4.12 & 0.14\\
4383.545 &  26.0 & 1.484 &      0.21 &  A & 4.32 & 0.14\\
4404.750 &  26.0 & 1.556 &     -0.15 &  A & 4.22 & 0.15\\
4415.123 &  26.0 & 1.607 &     -0.62 &  A & 4.27 & 0.15\\
4920.502 &  26.0 & 2.830 &      0.06 &   $-$ & 4.22 & 0.13\\
5269.537 &  26.0 & 0.858 &     -1.33 &  A & 4.27 & 0.16\\
5328.039 &  26.0 & 0.914 &     -1.47 &  A & 4.27 & 0.13\\
5371.489 &  26.0 & 0.957 &     -1.64 &  A & 4.32 & 0.16\\
5405.775 &  26.0 & 0.989 &     -1.85 &  A & 4.47 & 0.16\\
\hline
4923.922 &  26.1 & 2.891 &     -1.21 &  B & 4.12 & 0.01\\
5018.435 &  26.1 & 2.891 &     -1.35 & C & 4.27 & 0.01\\
5169.028 &  26.1 & 2.891 &     -0.87 &  B & 3.87 & 0.02\\
\hline
\end{tabular}
\end{table}
\begin{table}
\caption{Spectral line list - non-iron lines. The source of the atomic data is provided as Q, where D= Lawler and E= Sobeck.  NLTE corrections  ($\Delta$) are from the MPIA database.}
\label{tab:lines_others}
\begin{tabular}{ccccccc}
\hline
$\lambda$ &  Elem &  $\chi$ &  log(gf) &   Q & A(X) & $\Delta$ \\
(\AA) & & (eV) & & & LTE & NLTE \\
\hline
3829.355 &  12.0 & 2.707 &     -0.23 &    $-$ & 4.50 & 0.11   \\
3832.304 &  12.0 & 2.710 &      0.12 &    $-$ & 4.55 & 0.09   \\
5172.684 &  12.0 & 2.710 &     -0.40 &    $-$ & 4.40 & 0.12   \\
5183.604 &  12.0 & 2.715 &     -0.18 &    $-$ & 4.35 & 0.11   \\
3961.520 &  13.0 & 0.014 &     -0.32 &    $-$ & 2.6 & 0.50  \\
3905.523 &  14.0 & 1.907 &     -1.09 &    $-$ & 4.51 & 0.10  \\
4226.728 &  20.0 & 0.000 &      0.24 &    $-$ & 3.14 & 0.28 \\
4246.822 &  21.1 & 0.315 &      0.32 &    $-$ & 0.15 & $-$ \\ 
3913.461 &  22.1 & 1.115 &     -0.36 & D & 2.40 & 0.07  \\
4300.042 &  22.1 & 1.179 &     -0.46 & D & 2.25 & 0.06  \\
4395.031 &  22.1 & 1.083 &     -0.54 & D & 2.25 & 0.06  \\
4468.493 &  22.1 & 1.130 &     -0.63 & D & 2.15 & 0.01  \\
4501.270 &  22.1 & 1.115 &     -0.77 & D & 2.30 & 0.09  \\
4533.969 &  22.1 & 1.236 &     -0.77 &    $-$ & 2.40 & 0.09  \\
4571.971 &  22.1 & 1.571 &     -0.31 & D & 2.30 & 0.06  \\
4254.352 &  24.0 & 0.000 &     -0.09 & E & 2.44 & 0.56  \\
4274.812 &  24.0 & 0.000 &     -0.22 & E & 2.49 & 0.56  \\
4030.746 &  25.0 & 0.000 &     -0.50 &    $-$ & 2.18 & 0.62  \\ 
3858.297 &  28.0 & 0.422 &     -0.96 & D & 3.27 & $-$  \\
\hline

\end{tabular}
\end{table}


\bsp	
\label{lastpage}
\end{document}